\documentclass[12pt]{iopart}
\usepackage{iopams}

\usepackage{graphicx} 
\graphicspath{{figures/}}

\usepackage{amssymb,bm,dsfont,xcolor}
\usepackage{appendix}

\usepackage[colorlinks=true,linkcolor=blue,urlcolor=blue,citecolor=blue]{hyperref}

\usepackage[noabbrev]{cleveref}

\newcommand{\abs}[1]{|#1|}
\newcommand{\ket}[1]{|#1\rangle}
\newcommand{\bra}[1]{\langle#1|}
\newcommand{\braket}[2]{\langle#1|#2\rangle}

\begin{document}
	
	\title[Photo-assisted spin transport in double quantum dots with spin-orbit interaction]{Photo-assisted spin transport in double quantum dots with spin-orbit interaction}
	
	\author{David Fern\'andez-Fern\'andez$^1$, Jordi Pic\'o-Cort\'es$^{1, 2}$, Sergio Vela Liñ\'an$^3$ and Gloria Platero$^1$}
	
	\address{$^1$ Instituto de Ciencia de Materiales de Madrid ICMM-CSIC, 28049 Madrid, Spain}
	\address{$^2$ Institute for Theoretical Physics, University of Regensburg, 93040 Regensburg, Germany}
	\address{$^3$ Universidad Complutense de Madrid, Facultad de Ciencias Físicas, 28040 Madrid, Spain}
	\ead{david.fernandez@csic.es}
	\vspace{10pt}
	\begin{indented}
		\item[]\today
	\end{indented}
	
	\begin{abstract}
		We investigate the effect of spin-orbit interaction on the intra- and interdot particle dynamics of a double quantum dot under ac electric fields.
		The former is modeled as an effective ac magnetic field that produces electric-dipole spin resonance transitions, while the latter is introduced via spin-flip tunneling amplitudes.
		We observe the appearance of non-trivial spin-polarized dark states, arising from an ac-induced interference between photo-assisted spin-conserving and spin-flip tunneling processes.
		These dark states can be employed to precisely measure the spin-orbit coupling in quantum dot systems.
		Furthermore, we show that the interplay between photo-assisted transitions and spin-flip tunneling enables the system to operate as a highly tunable spin filter.
		Finally, we investigate the operation of the system as a resonant flopping-mode qubit for arbitrary ac voltage amplitudes, allowing for high tunability and enhanced qubit control possibilities.
	\end{abstract}
	
	\vspace{2pc}
	\noindent{\it Keywords}: semiconductor quantum dots, spin-orbit coupling, quantum transport, spin qubits, dark states, ac-driving dynamics and transport, electric dipole spin resonance
	
	\section{Introduction}
	Semiconductor spin qubits are among the most promising platforms for quantum computing \cite{Bogan2017,Vukusic2018,Hendrickx2020a,Hendrickx2021,Riggelen2022}, thanks to their long coherence times and high scalability potential \cite{Kloeffel2013,Veldhorst2017,Langrock2022}.
	In these systems, qubits are encoded in the spin states of electrons or holes localized in quantum dots (QDs).
	
	Manipulation of the qubit states is achieved by applying an oscillating magnetic field through electron spin resonance (ESR) \cite{Koppens2006,Veldhorst2015,Zwerver2022}.
	However, the localization of the required ac magnetic fields to address individual dots is challenging \cite{Koppens2006}.
	To overcome this limitation, an effective ac magnetic field is generated by electrically driving the particle in a material with strong spin-orbit coupling (SOC) \cite{Szumniak2012,Hendrickx2020,Mutter2021a,Jirovec2022,Wang2022,FernandezFernandez2022}, a magnetic field gradient \cite{Tokura2006,PioroLadriere2008,Watson2018}, a spatially dependent hyperfine field \cite{Laird2007,Ribeiro2010}, or modulated anisotropies of the effective $g$-factor \cite{Crippa2018,Hofmann2019,Liles2021}.
	These techniques enable electrical manipulation via the electric-dipole spin resonance (EDSR), which has been employed to obtain high-fidelity one- and two-qubit gates \cite{Zajac2018,Noiri2022,Vahapoglu2022,Mills2022,Xue2022}.
	
	In a double quantum dot (DQD), the particle (electron or hole) is delocalized between the two sites, giving rise to molecular-like orbitals.
	When an ac electric or magnetic field is applied, the resulting quantum coherence can lead to unusual properties, such as dynamical localization of charge and spin \cite{Dunlap1986,Grossmann1991,Aguado1997,Barata2000,GomezLeon2011}, spin filtering \cite{Hanson2004,Cota2005,Sanchez2006}, and long-range transfer mediated by photo-assisted tunneling (PAT) \cite{GallegoMarcos2015,Stano2015,PicoCortes2019}.
	These effects can be analyzed using Floquet theory \cite{Grifoni1998,Platero2004,Eckardt2015}, a mathematical framework used to study the behavior of periodically driven quantum systems.
	The time-dependent Schrödinger equation for the system is transformed into a time-independent problem by introducing a set of Floquet states that describe the system's behavior over one period of the drive.
	By analyzing the properties of these states, Floquet theory provides a powerful tool for understanding the steady-state behavior, response to external fields, and transport properties of driven quantum systems.
	Although these phenomena have been widely studied both theoretically and experimentally, the consequences of adding a strong SOC have only begun to be extensively analyzed in recent years \cite{Sala2021,Zhou2021,Zhou2022,Khomitsky2022}.
	In particular, the transport signatures under both strong SOC and large ac voltage amplitudes have not been widely investigated.
	
	Transport in this system is characterized by a spin-polarized current that arises from the finite probability of spin-flip tunneling between states of opposite spin in different QDs.
	When an ac electric field is present, these transitions can occur through the absorption or emission of one or more photons.
	We analyze the current and spin polarization through the system, including the effect of excited states via an effective magnetic field.
	Moreover, we find a set of non-trivial dark states (DS) in a particular parameter configuration where the interference between PAT spin-conserving and spin-flip transitions occurs.
	These processes give rise to a complex and nonlinear current output with remarkable features, including a potential read-out mechanism of the SOC.
	Finally, we explore the possibility of using the setup as a flopping-mode qubit, which allows for fully coherent manipulation of the spin through virtual transitions between the QDs in the presence of a strong SOC.
	The flopping-mode qubit has recently gained attention as a promising platform for quantum computing in solid-state systems~\cite{Croot2020,Mutter2021,Yu2022,Teske2023}.
	
	\section{Theoretical framework}\label{sec:theo-frame}
	
	\begin{figure}[!t]
		\centering
		\includegraphics{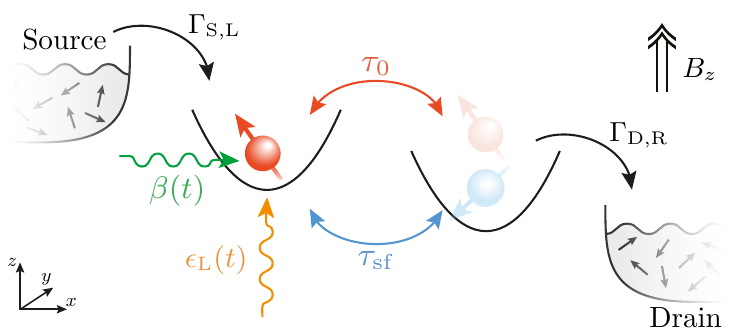}
		\caption{Schematic of a DQD system in the $x$-$y$ plane, with unpolarized contacts on the left and right sides.
			Particles can enter from the left lead with rate $\Gamma_\mathrm{S, L}$, and tunnel out to the right lead with rate $\Gamma_\mathrm{D,R}$.
			The system is subjected to a uniform magnetic field applied along the $z$-direction.
			The particle can tunnel between QDs following a spin-conserving path $\tau_0$.
			In addition to this path, a spin-flip tunneling $\tau_\mathrm{sf}$ is present due to the SOC.
			Finally, an ac electric field is applied to the left gate, denoted as $\epsilon_\mathrm{L}(t)$.
			In combination with the SOC, this gives rise to the OME term $\beta(t)$ (see text for details).}
		\label{fig:system_schematic}
	\end{figure}
	In this work, we aim to investigate the interference mechanisms between tunneling paths in a driven QD array.
	However, in QD arrays with three or more sites, additional features such as long-range transfer between distant QDs emerge.
	To focus on the primary consequences of adding an ac field in a system with strong spin-orbit coupling (SOC), we restrict our attention to the minimal model of DQD.
    This represents the smallest non-trivial scenario where interferences between different photoassisted interdot tunneling paths can be observed.
	
	The time-dependent Hamiltonian of a DQD system reads as ($\hbar = 1$)
	\begin{eqnarray}
		\label{eq:H_total}
		\hat{H}\left(t\right) =& \hat{H}_0(t)+ \hat{H}_1(t),\\
		\label{eq:H_0}
		\hat{H}_0(t) =& \sum_{\eta;\sigma}\epsilon_\eta(t)\hat{d}_{\eta,\sigma}^\dagger\hat{d}_{\eta,\sigma} + \sum_\eta \frac{E_{z}}{2}\hat{\sigma}_{z,\eta} \\
		\label{eq:H_1}&+\sum_{\eta\neq\eta';\sigma,\sigma'}\tau_{\eta,\sigma;\eta'\sigma'}\left(\hat{d}_{\eta,\sigma}^\dagger\hat{d}_{\eta',\sigma'} + h.c.\right), \nonumber \\
		\hat{H}_1(t) =&\sum_\eta \frac{\beta_\eta(t)}{2} \hat{\sigma}_{x,\eta},
		\label{eq:Hamiltonian_OME}
	\end{eqnarray}
	where $\hat{d}_{\eta, \sigma}$ ($\hat{d}_{\eta, \sigma}^\dagger$) is the annihilation (creation) operator at site $\eta \in \left\{\mathrm{L, R}\right\}$, with spin $\sigma\in\left\{\uparrow,\downarrow\right\}$.
	The first term in $\hat{H}_0$ represents the energy levels of the QDs, $\epsilon_\eta(t)$, which can be controlled by electric gates applied to individual dots.
	In particular, we consider a sinusoidal time-dependent gate in the leftmost dot, given by $\epsilon_\eta(t) = \epsilon_{\eta,0}+\epsilon_\mathrm{ac}\cos(\omega t)\delta_{\eta,\mathrm{L}}$.
	The second term is the Zeeman splitting due to the external magnetic field $B$ applied perpendicularly to the QDs plane, where $E_z = g\mu_B B_z$.
	The third term represents the hopping between the dots, and we consider two possible tunneling paths: a spin-conserving path $\propto \delta_{\sigma,\sigma}$ of amplitude $\tau_0$, and a spin-flip path with $\sigma \neq \sigma'$ of amplitude $\tau_\mathrm{sf}$ that arises due to the SOC, see \cref{fig:system_schematic}.
	The direction of the SOC determines the form of the spin-flip tunneling term, which is $\propto \bm{\alpha}\cdot\hat{\bm{\sigma}}$ where $\bm{\alpha}$ is the SOC vector~\cite{Burkard2002,Mutter2021}.
	Without loss of generality, we consider $\bm{\alpha}$ in the $y$-direction, such that $\tau_\mathrm{sf}\in \mathds{R}$ (see \ref{sec:OME_term}).
	Note that we neglect the Coulomb interaction term, as we focus on the single-particle dynamics.
	
	The Hamiltonian above does not explicitly include the excited states of the QD confining potential.
	However, when combined with the SOC and an ac electric field, the intra-dot dynamics leads to the emergence of an effective magnetic field~\cite{Rashba2008}, which is utilized for manipulating spin qubits in EDSR protocols.
	This effective magnetic field is contained in $\hat{H}_1(t)$ in our model.
	We obtain this term through a Schrieffer-Wolff transformation (SWT), and we present its derivation in \ref{sec:OME_term}.
	Under SOC, both the inter- and intra-dot dynamics contribute to the spin motion in distinct ways~\cite{Khomitsky2009,Khomitsky2011,Khomitsky2012}.
	Thus, we refer to this term as the orbital magneto-electric effect (OME) to distinguish it from the similar effective magnetic field that arises from the inter-dot dynamics of the spin, which we refer to as the tunneling magneto-electric effect (TME).
	This effect appears as a result of a finite spin-flip tunneling amplitude, and we discuss it in more detail in \cref{sec:Out-of-res}.
	
	The OME is orthogonal to the SOC vector, so it is oriented along the $x$-direction.
	Furthermore, if the electric field is homogeneous along the axis of the DQD, the OME term is the same in both dots, denoted by $\beta_\mathrm{L}(t)=\beta_\mathrm{R}(t)=\beta(t)$.
	Here, we consider $\beta(t)=\beta_\mathrm{SO}\cos(\omega t)$, where $\beta_{\mathrm{SO}}\in \mathds{R}$.
	Note that the frequency of the OME term is identical to that of $\epsilon_L(t)$ since it arises from the same applied voltage.
	We neglect the negligible rotation of the spin axis produced by the static contribution of the electric field in the following (see \ref{sec:OME_term}).
	Finally, we note that the OME term only appears for $E_z\neq 0$ since it requires breaking time-reversal symmetry.
	In the case where $E_z$ is small, the hyperfine interaction may produce an equivalent effect~\cite{Rashba2008}.
	
	The total Hamiltonian, written on the basis $\left\{\ket{L\uparrow}, \ket{L\downarrow}, \ket{R\uparrow},\ket{R\downarrow}\right\}$, reads
	\begin{equation}	
		\hat{H}=\left({\begin{array}{cccc}
				\epsilon_L(t)+E_{z}/2 & \beta(t) / 2 & -\tau_0 & -\tau_{\mathrm{sf}} \\
				\beta(t) / 2 & \epsilon_L(t)-E_{z}/2 & 	\tau_{\mathrm{sf}} & -\tau_0\\
				-\tau_0 & \tau_{\mathrm{sf}} & \epsilon_R + E_{z}/2 & \beta(t) /2 \\
				-\tau_{\mathrm{sf}} & -\tau_0 & \beta(t)/ 2 & \epsilon_R-E_{z}/2
		\end{array}}\right).
		\label{eq:total_hamiltonian}
	\end{equation}
	
	We introduce the parameter $\chi$, which characterizes the relationship between the spin-conserving and spin-flip tunneling amplitudes, as follows
	\begin{equation}
		\chi\equiv \frac{1}{\tau_0/\tau_{\mathrm{sf}}+1}.
		\label{eq:chi_SOC}
	\end{equation}
	In this expression, $\chi=0$ corresponds to $\tau_{\mathrm{sf}}=0$, while $\chi=1$ corresponds to $\tau_0=0$.
	We normalize the tunneling rates so that $\tau_0+\tau_\mathrm{sf}=\tau$.
	Most systems exhibit a much smaller spin-flip contribution to tunneling compared to the spin-conserving one~\cite{Maisi2016}.
	However, in certain cases, an applied external field can be used to tune the SOC and therefore the spin-flip contribution, for instance in GaAs-based hole QDs~\cite{Bogan2018,Bogan2021}.
	Additionally, synthetic SOC can be generated in DQDs via spin polarization in the external leads~\cite{Rohrmeier2021}.
	Hence, we consider arbitrary values of $\chi$ in the following.
	
	To read out the current, we couple the QD chain to source (S) and drain (D) leads, where the coupling is described by
	\begin{equation}
		\hat{H}_\Gamma = \sum_{l,\bm{k},\sigma,\eta}(\gamma_{l,\eta}\hat{c}_{l,\bm{k},\sigma}^\dagger\hat{d}_{\eta,\sigma} + h.c.),
	\end{equation}
	and $\hat{c}_{l,\bm{k},\sigma}$ is the annihilation operator for a particle in the lead $l=\left\{S,D\right\}$, with spin $\sigma$ and momentum $\bm{k}$.
	We assume the coupling between the leads and the QD chain to be spin-conserving.
	We further consider the infinite bias limit, in which transport is unidirectional from source to drain, and all the sidebands couple equally to the source lead.
	Hence, using the property of Bessel functions $\sum_n J_n(\epsilon_{\mathrm{ac}}/\omega)^2 =1$, we neglect the effect of the ac electric field on the renormalization of $\gamma_{l,\eta}$~\cite{GallegoMarcos2015}.
	
	We investigate the dynamics of the system by analyzing its reduced density matrix $\hat{\rho}(t)$, which can be described through the master equation under Markovian approximation
	\begin{equation}
		i \frac{d}{dt}\hat{\rho}(t) = [\hat{H}(t), \hat{\rho}(t)]+\mathcal{K}\hat{\rho}(t),
	\end{equation}
	where $\mathcal{K}\hat{\rho}$ is the kernel superoperator for weak coupling in the infinite bias approximation, and is defined as
	\begin{eqnarray}
		\mathcal{K}\hat{\rho} = \sum_{\sigma} &\left[\Gamma_{\mathrm{S,L}}\left(\hat{d}_{L,\sigma}^\dagger\hat{\rho}\hat{d}_{L,\sigma} - \frac{1}{2}\left\{\hat{d}_{L,\sigma}\hat{d}_{L,\sigma}^\dagger,\hat{\rho}\right\} \right)\right.
		\nonumber\\
		&\left.+\Gamma_{\mathrm{D,R}}\left(\hat{d}_{R,\sigma}\hat{\rho}\hat{d}_{R,\sigma}^\dagger - \frac{1}{2}\left\{\hat{d}_{R,\sigma}^\dagger\hat{d}_{R,\sigma},\hat{\rho}\right\} \right) \right].
	\end{eqnarray}
	Here, $\Gamma_{l,\eta}$ represents the transition rates due to coupling with leads.
	These are calculated as $\Gamma_{l,\eta} = 2\pi/|\gamma_{l,\eta}|^2D_l(\epsilon_F)$, with $D_l(\epsilon)$ being the density of states of lead $l$ and $\epsilon_F$ representing the Fermi energy.
	Our analysis focuses on strongly interacting QDs, where the energy difference between the single- and double-occupied states is much larger than the rest of the energy scales of the system.
	This allows us to investigate charge transport in the single-charge section of the stability diagram.
	
	The steady-state current can be calculated as follows
	\begin{equation}
		I^\infty = e\sum_\sigma \Gamma_{\mathrm{D,R}}\rho_{\mathrm{R}\sigma}^\infty,
	\end{equation}
	where $\rho_{\mathrm{R}\sigma}^\infty$ denotes the occupation of the rightmost dot with spin $\sigma$ in the stationary state, averaged over one period $T=2\pi/\omega$ of the ac voltage.
	More formally, this is given by
	\begin{equation}
		\rho_{\mathrm{R}\sigma}^\infty=\lim_{t\rightarrow\infty}\frac{1}{T}\int_t^{t+T} ds\bra{\mathrm{R}\sigma} \hat{\rho}(s)\ket{\mathrm{R}\sigma}.
	\end{equation}
	Finally, assuming an identical coupling to both leads, we set $\Gamma_\mathrm{S,L}=\Gamma_\mathrm{D,R}=\Gamma$.
	
	\section{Quantum transport in DQD with spin-orbit interaction}\label{sec:DQD}
	When the electric field is close to zero, quantum transport occurs predominantly at zero detuning $\delta\equiv\epsilon_{R,0}-\epsilon_{L,0}=0$, where the energy levels of the two dots are aligned.
	However, the presence of an ac field changes this picture by allowing for PAT, where a particle can tunnel from one dot to the other by absorbing or emitting a certain number of photons.
	An $n-$photon resonance occurs for $\delta = n\omega$, with $n\in\mathds{Z}$.
	
	\begin{figure}[!t]
		\centering
		\includegraphics{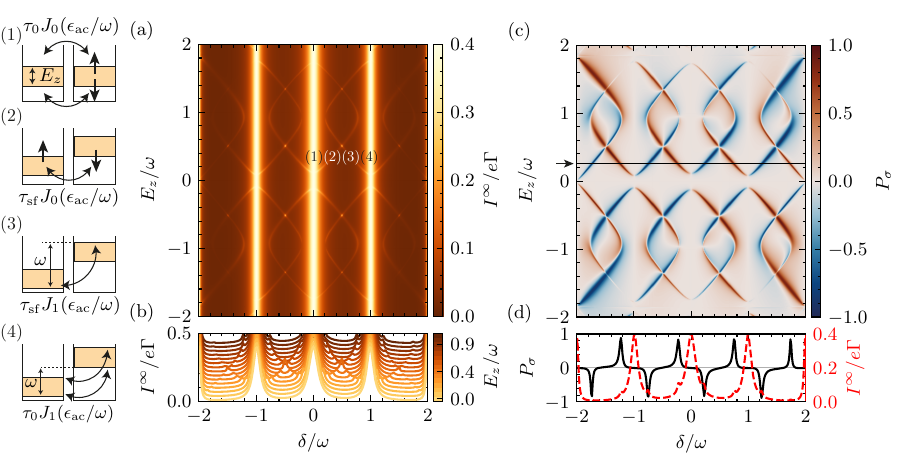}
		\caption{Panel (a) shows the current through a DQD as a function of $\delta$ and $E_z$ for $\chi=0.1$ and $\beta_{\mathrm{SO}}=E_z/2$.
			Four characteristic processes are highlighted in the figure and labeled as processes 1-4.
			These processes are represented on the left side of the figure.
			Process 1 corresponds to direct spin-conserving PAT through the two spin channels at $\delta=0$.
			Process 2 corresponds to direct spin-flip PAT, occurring when the energy splitting due to the magnetic field $E_z$ equals the detuning $\delta$.
			Process 3 corresponds to spin-flip PAT involving one photon, occurring when the energy splitting due to the magnetic field $E_z$ equals the detuning $\delta\pm\omega$.
			Process 4 corresponds to spin-conserving PAT involving one photon through the two spin channels at $\delta=\omega$.
			Panel (b) shows a set of cuts at different values of $E_z$ from panel (a) with a small offset of the $y$-axis for clarity.
			Panel (c) shows the spin polarization of the current as a function of $\delta$ and $E_z$.
			Panel (d) shows the spin polarization (solid black line, left axis) and total current (dashed red line, right axis) for $E_z=0.25\omega$ as a function of $\delta$, represented by the horizontal cut in panel (c).
			Other parameters are $\omega=10\tau=100\Gamma$, and $\epsilon_\mathrm{ac}=1.2\omega$.}
		\label{fig:current_magnetic_field}
	\end{figure}
	
	When $E_z\neq0$, the degeneracy of the spin doublets is broken, resulting in a current that is in general spin polarized.
	Several processes can be distinguished in this case, as schematized in \cref{fig:current_magnetic_field}.
	First, there are spin-conserving resonances with either no photons (process 1), known as direct resonances, or with the emission/absorption of a photon (process 4).
	These resonances occur at $\delta=n\omega$, where the particle can tunnel directly from one dot to the other.
	These resonances survive in the presence of a magnetic field, since the difference in energy between a particle with the same spin in different dots is also $\delta$.
	Moreover, since the levels are now split in spin by the magnetic field, there can also be resonances when states with opposite spin have the same energy, which is enabled by the presence of a spin-flip component in the tunneling amplitudes (process 2).
	This occurs whenever $\delta=\pm E_z$.
	In the presence of an ac bias, these resonances are accompanied by a set of replicas due to PAT (process 3).
	
	As a result, when the current is represented as a function of $\delta$ and $E_z$ as in \cref{fig:current_magnetic_field}(a-b), we observe both the usual resonances at $\delta=n\omega$ and a set of spin-flip resonances along the lines
	\begin{equation}
		\delta+n\omega = \pm E_z.
	\end{equation}
	In the $\delta-E_z$ representation of \cref{fig:current_magnetic_field}(a), these correspond to diagonal lines.
	Note that transport at these resonances is not blocked when a particle tunnels from the source into the left dot with a direction that is not energy-aligned with any state on the right dot (such as $\ket{L\downarrow}$ in process 2 of \cref{fig:current_magnetic_field}).
	This is because the spin states always have a small component of the opposite spin due to the finite spin-flip tunneling amplitude.
	Alternatively, we can view the spin rotation as a result of virtual tunneling to the dot via the TME, which causes the spin to align with the resonance.
	
	While the basic picture described earlier provides a useful starting point, there are several deviations from it that must be taken into account.
	One of these occurs in the vicinity of a PAT process involving $n$ photons.
	In this regime, the states in the two dots become strongly hybridized with energies
	\begin{equation}
		E_\pm = \delta\pm \tau J_n\left(\frac{\epsilon_\mathrm{ac}}{\omega}\right),
		\label{eq:hybrid-energies}
	\end{equation}
	where $J_n(z)$ is the $n$th-order Bessel function.
	As a result, the diagonal lines are no longer straight but curve near $E_z\approx\pm\tau$ due to the resonance conditions with these energies.
	This effect is clearly visible in \cref{fig:current_magnetic_field}(a) for the PAT resonances near $E_z=0$, with the overall shape of these resonances closely following hyperbolas in the $\delta-E_z$ representation.
	
	Another deviation from the simple picture occurs near $E_z=\pm\omega$, where the OME field induces resonant transitions between the two spin states.
	In this regime, the spin-flip resonances do not intersect the spin-conserving PAT resonances at $\delta=\pm\omega$ but are instead repelled, creating a current-free gap between the main resonances (process 1) and the spin-flip resonances (process 2).
	Additionally, at the main resonances where $\delta=0$, the spin-flip resonances cross at $E_z=\pm(\tau+\beta_{\mathrm{SO}})$ instead of simply crossing at $E_z=\pm \tau$, as discussed earlier.
	The spin-flip resonances remain hyperbolic but are vertically displaced from where they would be for $\beta_{\mathrm{SO}}=0$.
	
	Finally, when photo-assisted spin-flip resonances involving zero or one photon meet, they do not become distorted (as at $\delta=\pm \omega/2$, $E_z=\pm \omega/2$ in \cref{fig:current_magnetic_field}(a)) since they correspond to processes that transmit different spin polarizations.
	
	The spin polarization of the current is given by the expression
	\begin{equation}
		P_\sigma = \frac{I_\uparrow^\infty - I_\downarrow^\infty}{I_\uparrow^\infty + I_\downarrow^\infty},
	\end{equation}
	where $I_\sigma^\infty = e\Gamma\rho_{\mathrm{R},\sigma}^\infty$.
	In \cref{fig:current_magnetic_field}(c), we present $P_\sigma$ as a function of $\delta$ and $E_z$.
	As expected, the spin polarization is non-zero at the spin-flip current branches, provided that only one path is active (either $\ket{L\uparrow}\rightarrow\ket{R\downarrow}$ or $\ket{L\downarrow}\rightarrow\ket{R\uparrow}$).
	We can achieve a spin polarization close to 1, with the largest limitation being the overlap of a spin-flip resonance with a spin-conserving PAT resonance.
	The width of spin-conserving resonances can be tuned by reducing $\tau_0$, and the separation between PATs can be increased by varying the ac voltage frequency $\omega$.
	
	This setup offers several ways to control spin polarization.
	In \cref{fig:current_magnetic_field}(d), we show the polarization along with the total current.
	By varying $\delta$ from $\delta\simeq 0.2\omega$ to $\delta\simeq0.8\omega$, we can shift from a direct PAT resonance with a strong spin polarization in one direction to a one-photon PAT resonance with a strong spin polarization in the opposite direction.
	Hence, polarization can be inverted without crossing the resonance at $\delta=0$.
	Since $\delta$ is often one of the easiest parameters to control in experimental setups, this allows us to generate highly tunable fully spin-polarized currents.
	Remarkably, this can be achieved even when the spin-flip amplitude is considered very small ($\chi=0.1$).
	With a correct tuning of $\chi$ and $\tau$ (reducing the width of the spin-conserving resonances), the spin polarization can be adjusted to an even larger degree.
	
	We note a slight asymmetry in the spin polarization in the two adjacent highly polarized peaks at $\delta\simeq 0.2\omega$ and $\delta\simeq0.8\omega$, resulting from the OME term.
	This is evident in \cref{fig:current_magnetic_field}(c), where the resonance lines have different widths depending on whether they cross the $E_z=\pm\omega$, $\delta=\pm\omega$ points or not.
	For $\beta_{\mathrm{SO}}\rightarrow0$, the current associated with the two resonances has the same absolute value of polarization (and is close to $P_\sigma=\pm 1$, respectively).
	
	Tuning the frequency of the ac voltage allows for another method of generating spin-polarized currents in the setup.
	The DQD is initialized in a configuration where direct resonances without photons are energetically disfavored, and the voltage frequency is tuned to either the spin-conserving or the spin-flip resonance with absorption or emission of photons.
	The resulting current is spin-polarized, with the polarization defined by spin state in resonance.
	This enables fully electric control of the spin polarization of the current without the need to modify the magnetic fields, allowing for fast control of the current under experimental conditions.
	
	The effect of varying the amplitude of the ac voltage is then studied.
	Due to Landau-Zener-Stückelberg (LZS) interferences,  if the QDs are in resonance at $\delta=n\omega$, then photo-assisted transitions occur with a tunneling amplitude that is renormalized by a Bessel function as~\cite{Grifoni1998,Platero2004,KOHLER2005}
	\begin{equation}
		\tau_\zeta\rightarrow \tau_\zeta J_n\left(\frac{\epsilon_\mathrm{ac}}{\omega}\right), \quad \zeta\in\{0, \mathrm{sf}\}.
		\label{eq:renormalized_tunneling}
	\end{equation}

	\begin{figure}
		\centering
		\includegraphics{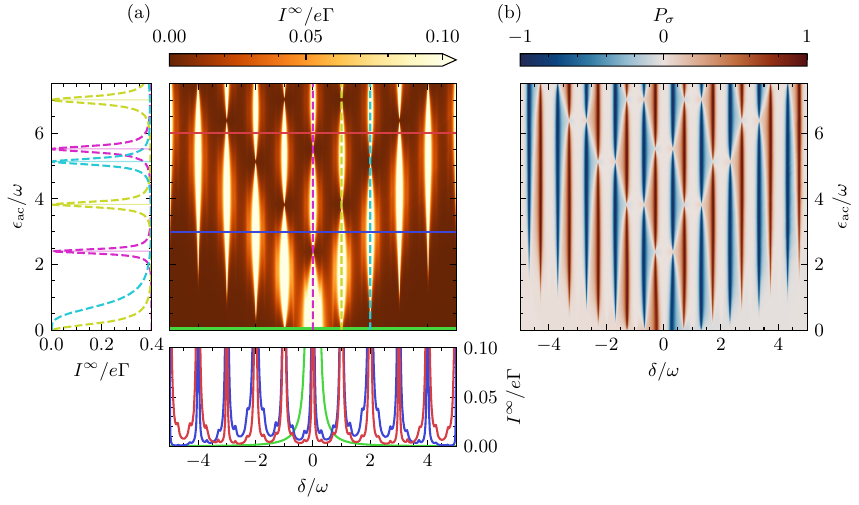}
		\caption{Current (a) and spin polarization (b) through a DQD as a function of $\delta$ and $\epsilon_\mathrm{ac}$.
			In panel (a, bottom), we observe the appearance of satellite peaks close to the main resonances as the ac voltage amplitude increases from zero (green) to $3\omega$ (blue) and $6\omega$ (red).
			Additionally, in panel (a, left) we observe coherent destruction of tunneling at certain values of $\delta=n\omega$ with $n=0$ (pink), $n=1$ (lime), and $n=2$ (cyan).
			The zeros of the Bessel function $J_n(\epsilon_\mathrm{ac}/\omega)$ are denoted by the horizontal solid lines.
			The parameters used are $\omega=10\tau=100\Gamma$, $E_z=0.3\omega$, $\beta_{\mathrm{SO}}=0$, and $\chi = 0.2$.}
		\label{fig:LZS}
	\end{figure}

	Crucially, both the spin-conserving and spin-flip amplitudes are modified by the presence of the ac field~\footnote{Note that this renormalization was already taken into account in \cref{eq:hybrid-energies}.}.
	This expression holds except at the points where the spin levels are in resonance with $E_z = m\omega$, as we will see in the next sections.
	
	The described system exhibits coherent destruction of tunneling (CDT) due to the presence of the ac field.
	The CDT effect occurs when both spin-conserving and spin-flip tunneling rates are suppressed due to destructive interference at the zeros of a given Bessel function.
	This effect is visible in \cref{fig:LZS}(a), which displays the current as a function of both $\delta$ and $\epsilon_\mathrm{ac}$, where the main resonances are seen with thick parallel lines along the $\epsilon_\mathrm{ac}$ axis.
	Spin-flip resonances are visible as two thinner lines at each side of the main resonances.
	Cuts at several values of $\epsilon_\mathrm{ac}$ are shown in the bottom panel of \cref{fig:LZS}(a), while the interference pattern leading to CDT is shown in the left panel of the same figure.
	The solid lines in the left panel of \cref{fig:LZS}(a) denote the first zeros of $J_n(\epsilon_\mathrm{ac}/\omega)$, which coincide with the dips in the current at $\delta=n\omega$.
	In addition, \cref{fig:LZS}(b) displays the spin polarization for the LZS interferometry, where the main spin-conserving resonances have an unpolarized spin current $P_\sigma\sim 0$, since both spin-up and spin-down channels are present with the same probability.
	On the other hand, the spin-flip resonances at $\delta=n\omega \pm E_z$ are highly spin polarized $P_\sigma \sim \pm 1$.
	It is clear from this figure that spin-flip resonances also exhibit the CDT effect where $J_n(\epsilon_\mathrm{ac}/\omega) = 0$.
	
	\subsection{Effect of the OME}\label{sec:DQD_global_OME}
	In this section, we will explore the impact of the OME term on the system dynamics.
	Although the amplitude of the OME field induced by SOC is typically weaker than that of the electric voltage, it can still have a significant effect near the spin resonances at $E_z\simeq n\omega$.
	In this regime, the effective ac magnetic field can induce resonant transitions between the two spin states on the same QD.
	To examine this effect more closely, we consider the Hamiltonian at any such resonance and perform the unitary transformation given by
	\begin{equation}
		\hat{U}(t)=\exp\left(\frac{-in\omega t \hat{\sigma}_z}{2}\right)\exp\left(\frac{-i\epsilon_\mathrm{ac}\sin(\omega t) (1 + \hat{\tau}_z)}{2\omega}\right),
		\label{eq:Ut}
	\end{equation}
	where $\hat{\tau}_i$ are Pauli matrices associated to the charge (left/right dot) degree of freedom.
	The leftmost operator takes the system into the rotating frame at the spin resonance, whereas the rightmost operator transforms the system into the interacting picture with respect to the ac voltage.
	The unitary transformation effectively removes the driving term from the Hamiltonian while turning the tunneling time-dependent.
	This transformation allows us to work within a rotating-wave approximation (RWA)\cite{Eckardt2015}, where we can consider arbitrary ac voltage amplitudes.
	
	After applying the above transformation and employing the Jacobi-Anger expansion, the Hamiltonian terms are modified as follows
	\begin{numparts}
		\begin{eqnarray}
			\bra{\eta\sigma}\hat{H}_0\ket{\eta\sigma}=E_{z}\to E_{z}-n\omega, \\
			\bra{\mathrm{R}\sigma}\hat{H}_0\ket{\mathrm{L}\sigma}=\tau_{0}\to\tau_{0}\sum_{k=-\infty}^{\infty}J_{k}\left(\frac{\epsilon_{\mathrm{ac}}}{\omega}\right)e^{ik\omega t},\\
			\bra{\mathrm{R}\uparrow(\downarrow)}\hat{H}_0\ket{\mathrm{L}\downarrow(\uparrow)}=\pm\tau_{\mathrm{sf}}\to\pm\tau_{\mathrm{sf}}\sum_{k=-\infty}^{\infty}J_{k}\left(\frac{\epsilon_{\mathrm{ac}}}{\omega}\right)e^{i\left(k\mp n\right)\omega t},\\
			\bra{\eta \uparrow}\hat{H}_1\ket{\eta \downarrow}=\beta\left(t\right)\to\frac{\beta_{\mathrm{SO}}}{2}\left(e^{i\left(n+1\right)\omega t}+e^{i\left(n-1\right)\omega t}\right).
		\end{eqnarray}
	\end{numparts}
	To obtain additional terms in the effective Hamiltonian, we impose hermiticity on the Hamiltonian.
	We now focus on the avoided crossing at $\delta\approx0$, $n= 1$, as shown in \cref{fig:current_magnetic_field}, where the resonance $n=-1$ follows a similar behavior.
	We apply the RWA, neglecting the time-dependent terms, which yields the effective Hamiltonian
	\begin{equation}
		\hat{H}_{\mathrm{RWA}}=\left(\begin{array}{cccc}
			-\delta/2 & \beta_{\mathrm{SO}}/4 & -\tau_\mathrm{0}^\mathrm{RWA} & -\tau_\mathrm{sf}^\mathrm{RWA}\\
			\beta_{\mathrm{SO}}/4 & -\delta/2 & -\tau_\mathrm{sf}^\mathrm{RWA} & -\tau_\mathrm{0}^\mathrm{RWA}\\
			-\tau_\mathrm{0}^\mathrm{RWA} & -\tau_\mathrm{sf}^\mathrm{RWA} & \delta/2 & \beta_{\mathrm{SO}}/4\\
			-\tau_\mathrm{sf}^\mathrm{RWA} & -\tau_\mathrm{0}^\mathrm{RWA} & \beta_{\mathrm{SO}}/4 & \delta/2
			\label{eq:HRWA}
		\end{array}\right).
	\end{equation}
	Here, $\tau_\mathrm{0}^\mathrm{RWA}=\tau_0J_0(\epsilon_\mathrm{ac}/\omega)$ and $\tau_\mathrm{sf}^\mathrm{RWA}=\tau_\mathrm{sf}J_1(\epsilon_\mathrm{ac}/\omega)$.
	Note that at the resonance between the two spin levels, the spin-conserving tunneling amplitude is renormalized by $J_0(z)$, while the spin-flip amplitude is renormalized by $J_1(z)$, with $z=\epsilon_\mathrm{ac} / \omega$.
	
\begin{figure}
		\centering
		\includegraphics{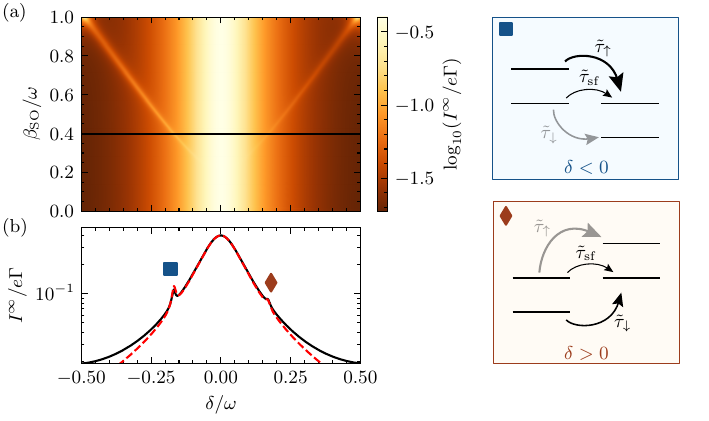}
		\caption{Panel (a) shows the current through a DQD as a function of $\delta$ and $\beta_\mathrm{SO}$.
			Panel (b) displays a horizontal cut of panel (a) at $\beta_\mathrm{SO}=0.4\omega$, where the black solid line corresponds to the current obtained from the original Hamiltonian given in \cref{eq:total_hamiltonian}, and the red dashed line corresponds to the current obtained from the effective Hamiltonian given in \cref{eq:HRWA_2}.
			The current exhibits two satellite peaks located at $\delta \simeq \pm \beta_\mathrm{SO}/2$, indicated by blue square and red diamond symbols.
			The energy level diagram on the right-hand side of the figure illustrates the energy levels and tunneling paths for the effective Hamiltonian.
			The width of the arrows denotes the tunneling amplitudes for each possible path, while the gray arrows represent tunneling paths that are less favorable.
			Throughout the panels, we set $E_z = \omega$, $\epsilon_{\mathrm{ac}}=1.2\omega$, $\omega=10\tau=100\Gamma$, and $\chi = 0.2$.}
		\label{fig:asymmetry}
	\end{figure}

	In this rotating frame, the OME term acts as a constant magnetic field, leading to a splitting $\beta_\mathrm{SO}/2$ of the spins in the $x$-direction.
	The avoided crossing between the spin-conserving and spin-flip resonances discussed above appears because the OME term produces a splitting between the spin states that is only strong near the EDSR.
	As a result, resonant transport occurs when $\delta = \pm \beta_\mathrm{SO}/2$.
	These new resonances appear as side peaks next to the main resonance at $\delta = 0$, as shown in \cref{fig:asymmetry}(a).
	We observe that the position of the resonances exhibits a linear relationship between detuning and $\beta_\mathrm{SO}$, as expected.
	However, for $\beta_\mathrm{SO}\lesssim\tau$, the hybridization between states is large enough to distort this simple picture, in a similar way to \cref{eq:hybrid-energies}.
	
	The above results hold true only when $\beta_\mathrm{SO}/\omega\ll1$, as required by the RWA.
	However, we can extend the validity of our analysis by applying the following transformation to the original Hamiltonian
	\begin{equation}
		\hat{U}'(t)=\exp\left(\frac{-i\beta_{\mathrm{SO}}\sin\left(\omega t\right)\hat{\sigma}_{x}}{2\omega}\right).
		\label{eq:Ut-prime}
	\end{equation}
	After this transformation, we apply $\hat{U}(t)$ from \cref{eq:Ut}.
	The total transformation eliminates the OME term, but transforms the spin-flip and Zeeman splitting terms as
	\numparts
	\begin{eqnarray}
		\frac{E_{z}}{2}\hat{\sigma}_{z}\to&\frac{E_{z}}{2}\cos\left[\frac{\beta_{\mathrm{SO}}}{\omega}\sin\left(\omega t\right)\right]\hat{\sigma}_{z}\nonumber\\
		&+\frac{E_{z}}{2}\sin\left[\frac{\beta_{\mathrm{SO}}}{\omega}\sin\left(\omega t\right)\right]\hat{\sigma}_{y},\\
		\tau_{\mathrm{sf}}\hat{\tau}_{y}\hat{\sigma}_{y}\to&
		\tau_{\mathrm{sf}}\hat{\tau}_y\cos\left[\frac{\beta_{\mathrm{SO}}}{\omega}\sin\left(\omega t\right)\right]\hat{\sigma}_{y}\nonumber\\
		\label{eq:OME_term_2}
		&-\tau_{\mathrm{sf}}\hat{\tau}_y\sin\left[\frac{\beta_{\mathrm{SO}}}{\omega}\sin\left(\omega t\right)\right]\hat{\sigma}_{z}.
	\end{eqnarray}
	\endnumparts
	The first term on the right-hand side of the expression for $E_z$ reflects the renormalization effect due to the ac magnetic field.
	This term modifies the Zeeman splitting by a factor $J_0(\beta_\mathrm{SO}/\omega)$ compared to the result in \cref{eq:HRWA}.
	When $J_0(\beta_\mathrm{SO}/\omega)=0$ due to destructive interference, the Zeeman splitting is completely suppressed in a process known as spin locking~\cite{GomezLeon2011}.
	However, this suppression occurs for values of $\beta_\mathrm{SO}$ beyond those discussed in \ref{sec:OME_term}.
	In general, the main effect of this renormalization is that EDSR now occurs at $E_z J_0(\beta_\mathrm{SO}/\omega)=n\omega$.
	
	The second term on the expression for $E_z$ reduces, for $\beta_\mathrm{SO}/\omega\ll1$, to the (phase-shifted) OME term.
	At resonance $n=\pm 1$, this corresponds to the substitution
	\begin{equation}
		\frac{\beta_\mathrm{SO}}{2}\to E_z J_1\left(\frac{\beta_\mathrm{SO}}{\omega}\right),\label{eq:beta_SO-RWA}
	\end{equation}
	as shown in equation (\ref{eq:HRWA}).
	In the limit $\beta_\mathrm{SO}/\omega\ll1$ we recover the results shown above.
	When $J_1(\beta_\mathrm{SO}/\omega)=0$, the effective magnetic field is suppressed in a similar way to the case of spin locking.
	However, this effect occurs for parameters beyond the scope of the approximations of \ref{sec:OME_term} and we do not consider it further.
	
	Regarding the transformed spin-flip term given in equation~(\ref{eq:OME_term_2}), it may seem that the first term would also produce a similar Bessel function renormalization.
	However, the spin-flip amplitude at an arbitrary $\beta_\mathrm{SO}$ allows sidebands to be generated from both the ac voltage and the OME term.
	The combination of these sidebands yields a more complicated expression.
	After applying the transformation of \cref{eq:Ut} and expanding the first term of equation~(\ref{eq:OME_term_2}) using the Jacobi-Anger expansion, we obtain the following expression
	\begin{equation}
		\fl
		\pm\tau_{\mathrm{sf}}\to\frac{\pm\tau_\mathrm{sf}}{2} \sum_{k=-\infty}^\infty \sum_{k'=-\infty}^\infty J_{k}\left(\frac{\epsilon_{\mathrm{ac}}}{\omega}\right)J_{2k'}\left(\frac{\beta_{\mathrm{SO}}}{\omega}\right)\left(e^{i(-k\mp n+2k')\omega t}+e^{i(-k\mp n-2k')\omega t}\right),
	\end{equation}
	where $\pm$ corresponds to the sign of the two spin channels, as in \cref{eq:total_hamiltonian}.
	In the RWA, we find
	\begin{eqnarray}
		\pm\tau_\mathrm{sf}\to& (\mp 1)^{n+1}\tau_\mathrm{sf}^\mathrm{RWA},\\
		\tau_\mathrm{sf}^\mathrm{RWA}=& \tau_\mathrm{sf}\sum_{k=-\infty}^\infty J_{2k + n}\left(\frac{\epsilon_\mathrm{ac}}{\omega}\right)J_{2k}\left(\frac{\beta_\mathrm{SO}}{\omega}\right).
		\label{eq:tau_sf_RWA}
	\end{eqnarray}
	
	Finally, consider the last term on equation~(\ref{eq:OME_term_2}).
	After applying the transformation of \cref{eq:Ut}, we find a term $\propto\ket{L}\bra{R}\hat{\sigma}_z$ with amplitude
	\begin{equation}
		\tau_{\mathrm{sf}}\sum_{k=-\infty}^{\infty}\sum_{k'=0}^{\infty}J_{k}\left(\frac{\epsilon_{\mathrm{ac}}}{\omega}\right)J_{2k'+1}\left(\frac{\beta_{\mathrm{SO}}}{\omega}\right)\left(e^{i\left(k+2k'+1\right)\omega t}-e^{i\left(k-2k'-1\right)\omega t}\right).
	\end{equation}
	In the RWA, only the terms with $k=\pm(2k'+1)$ contribute and we obtain the real tunneling rate 
	\begin{equation}
		\tau_\mathrm{si}^\mathrm{RWA}=\tau_{\mathrm{sf}}\sum_{k=-\infty}^{\infty}J_{2k+1}\left(\frac{\epsilon_{\mathrm{ac}}}{\omega}\right)J_{2k+1}\left(\frac{\beta_{\mathrm{SO}}}{\omega}\right).
		\label{eq:tau_si}
	\end{equation}
	This produces spin-conserving amplitudes of different absolute values depending on the direction of the spin, i.e., the spin-conserving tunneling of a spin-down particle occurs with a tunneling amplitude different from that of a spin-up particle.
	Note that this term arises only because of the combination of sidebands from both the ac voltage and the OME term, resulting in a finite static component that contributes to the Hamiltonian even in the RWA.
	Remarkably, both electric and magnetic sidebands arise from the same applied voltage.
	In that sense, this is a constructive self-interference effect, in a similar way to virtual tunneling processes that involve two PAT transitions~\cite{GallegoMarcos2015,PicoCortes2019}.
	
	Let us now compare the results for arbitrary $\beta_\mathrm{SO}$ with the Hamiltonian in the $\beta_\mathrm{SO}\ll\omega$ limit, \cref{eq:HRWA}.
	Again, let us focus on $n=1$, $\delta\approx 0$.
	We diagonalize the spin subspace by performing a rotation around the y-axis by an angle $\theta \equiv \arctan(E_z^\mathrm{RWA}/\beta_\mathrm{SO}^\mathrm{RWA})$, where $E_z^\mathrm{RWA}\equiv E_zJ_0\left(\beta_\mathrm{SO}/\omega\right) - \omega$ and $\beta_{\mathrm{SO}}^\mathrm{RWA} \equiv E_z J_1(\beta_\mathrm{SO}/\omega)$.
	Finally, the effective Hamiltonian in matrix form reads
	\begin{equation}
		\hat{\widetilde{H}}=\left(\begin{array}{cccc}
			(\widetilde{E}_z-\delta) / 2 & 0 & -\widetilde{\tau}_\uparrow & -\widetilde{\tau}_\mathrm{sf}\\
			0 & (-\widetilde{E}_z-\delta)/2 & -\widetilde{\tau}_\mathrm{sf} & -\widetilde{\tau}_\downarrow\\
			-\widetilde{\tau}_\uparrow & -\widetilde{\tau}_\mathrm{sf} & (\widetilde{E}_z+\delta)/2 & 0\\
			-\widetilde{\tau}_\mathrm{sf} & -\widetilde{\tau}_\downarrow & 0 & (-\widetilde{E}_z+\delta)/2
			\label{eq:HRWA_2}
		\end{array}\right),
	\end{equation}
	where we have defined the effective magnitudes as
	\begin{eqnarray}
		\widetilde{E}_z&\equiv \sqrt{(E_z^\mathrm{RWA})^2+(\beta_{\mathrm{SO}}^\mathrm{RWA})^2},\\
		\widetilde{\tau}_{\uparrow,\downarrow} &\equiv\tau_0^\mathrm{RWA} \mp \left[\sin(\theta)\tau_\mathrm{sf}^\mathrm{RWA} - \cos(\theta)\tau_\mathrm{si}^\mathrm{RWA}\right],\\
		\widetilde{\tau}_\mathrm{sf} &\equiv \cos(\theta)\tau_\mathrm{sf}^\mathrm{RWA} + \sin(\theta)\tau_\mathrm{si}^\mathrm{RWA}.
	\end{eqnarray}
	The effective model describes a DQD system subject to a homogeneous Zeeman splitting $\widetilde{E}_z$, where the tunneling rates for spin-up and spin-down channels are generally different, denoted as $\widetilde{\tau}_{\uparrow}$ and $\widetilde{\tau}_{\downarrow}$, respectively, and there is also a spin-flip tunneling rate $\widetilde{\tau}_\mathrm{sf}$.
	When the OME term $\beta_\mathrm{SO}$ is much smaller than the driving frequency, the effective magnetic field can be approximated as $\widetilde{E}_z = \beta_\mathrm{SO}/2+\mathcal{O}\left((\beta_\mathrm{SO}/\omega)^5\right)$, which gives a linear relationship between the detuning $\delta$ and $\beta_\mathrm{SO}$ for the resonant condition discussed earlier.
	Note that the effective model is only applicable for small detunings, i.e., $\delta\ll\omega$.
	
	The different tunneling rates $\widetilde{\tau}_{\uparrow,\downarrow}$ contribute in a characteristic way to the current, leading to a significant asymmetry between the two satellite peaks shown in \cref{fig:asymmetry}(b), depending on the sign of the detuning $\delta$.
	This asymmetry arises fundamentally due to the fact that the magnitudes of the tunneling rates for spin-up and spin-down channels are not equal, i.e., $\abs{\widetilde{\tau}_{\uparrow}}\neq\abs{\widetilde{\tau}_{\downarrow}}$.
	The two satellite peaks occur when the energy levels of the two dots with opposite spin are in resonance, as illustrated in \cref{fig:asymmetry}~\footnote{Note that the spin-up and spin-down levels in the effective model are a non-trivial combination of the spin-up and spin-down levels of our original model.
	However, the population of the left and right dots in the effective model of \cref{eq:HRWA_2} still resembles the same populations as in the original system of \cref{eq:total_hamiltonian}.}.
	If $\delta < 0$, the spin-down state of the left dot is in resonance with the spin-up state of the right dot, and the dominant tunneling process is through $\widetilde{\tau}_\mathrm{sf}$, while $\widetilde{\tau}_\downarrow$ can be neglected to first order.
	On the other hand, the spin-up energy level on the left dot is out of resonance with an energy difference given by $\widetilde{E}_z$, but a small but finite probability of tunneling to the right dot exists due to $\widetilde{\tau}_\uparrow$ \cite{Gurvitz1996}.
	In the case of $\delta>0$, a similar situation occurs, except that the non-resonant level (spin-down in this case) tunnels with an amplitude given by $\widetilde{\tau}_\downarrow$.
	Using equations~(\ref{eq:tau_sf_RWA}, \ref{eq:tau_si}) with $k= -2, -1, \ldots, 2$ and the parameters shown in \cref{fig:asymmetry}(b), we obtain $\widetilde{\tau}_\uparrow \simeq 0.06\omega$, $\widetilde{\tau}_\downarrow \simeq 0.04\omega$, and $\widetilde{\tau}_\mathrm{sf} \simeq 0.002 \omega$.
	Since $\widetilde{\tau}_\uparrow > \widetilde{\tau}_\downarrow$, the effective model predicts a higher current for the resonance at $\delta <0$, which is consistent with the numerical results obtained from the original Hamiltonian.
	In \cref{fig:asymmetry}(b), we compare the current obtained both with the total Hamiltonian of \cref{eq:total_hamiltonian} and with the effective Hamiltonian \cref{eq:HRWA_2}, showing how the effective Hamiltonian predicts the presence of satellite peaks close to the main resonance and the asymmetry in $\delta$.
	
	The fact that the asymmetry in $\delta$ is a consequence of $\abs{\widetilde{\tau}_{\uparrow}}\neq\abs{\widetilde{\tau}_{\downarrow}}$ can be further seen by calculating the effective magnetic field due to the TME.
	This is shown in more detail in \ref{sec:TME-spin-resonance}.
	
	\section{Dark state formation}
	Dark states (DSs) are a well-known feature in open quantum systems.
	They were first described in the context of the optical response of electrons to laser pumping.
	In mesoscopic transport, a DS refers to a steady state in which coherent interference results in a current blockade, even in a situation where current flow would be expected~\cite{Michaelis2006,Plotl2009,Busl2010,Donarini2019}.
	In this specific setup, DSs correspond to a particle being confined to the left dot, as the right dot would immediately empty through tunneling to the drain.
	
	We will focus on the main resonances $\delta = m\omega$, $m\in\mathbb{Z}$, where non-zero current flow is expected.
	In this case, as mentioned earlier, current blockade occurs due to the LZS interference pattern at the points where $J_m\left(\epsilon_\mathrm{ac}/\omega\right)=0$, yielding the aforementioned CDT.
	As described earlier, this renormalization is the same for both the spin-conserving and spin-flip tunneling amplitudes.
	However, when the system is simultaneously in a configuration in which the spin states are also in resonance with the ac voltage frequency $E_z=n\omega$, the renormalization is not necessarily the same, as seen in the previous section.
	We will focus on this case in the following.
	
\begin{figure}
		\centering
		\includegraphics{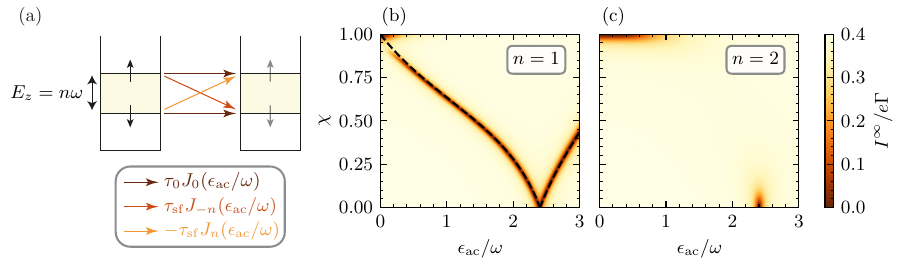}
		\caption{Panel (a) shows a scheme of a DQD system subjected to a magnetic field that induces a Zeeman splitting $E_z=n\omega$.
			The spin-conserving tunneling rate is renormalized as $\tau_0\rightarrow\tau_0 J_0(\epsilon_\mathrm{ac}/\omega)$, and the spin-flip rate becomes $\tau_{\mathrm{sf}}\rightarrow \tau_{\mathrm{sf}}J_{\pm n}(\epsilon_\mathrm{ac}/\omega)$.
			Panels (b-c) show the current through the DQD as a function of $\chi$ and $\epsilon_\mathrm{ac}$.
			The Zeeman splitting is $E_z=\omega$ (b) and $E_z=2\omega$ (c), and the detuning $\delta$ is set to zero.
			The dashed line corresponds to the analytical prediction for the DS condition $\widetilde{\chi}^{(n)}$, which only has real solutions for odd $n$ (see \cref{eq:condition_DS}).
			For $n=2$, the dark state condition cannot be satisfied.
			The parameters are $\omega=10\tau=100\Gamma$ and $\beta_\mathrm{SO}=0$.}
		\label{fig:DQD_dark_state_chi_eps}
	\end{figure}

	Let us consider the Hamiltonian in the rotating frame, as given in \cref{eq:HRWA}, for an $n-$photon resonance.
	First, let us assume $\beta_\mathrm{SO}=0$, we will analyze below how the dark states are affected when we include a non-zero OME term.
	In the case of a spin-up particle tunneling from the left to the right dot, emitting $n$ photons, the spin-flip tunneling rate is renormalized as $\tau_\mathrm{sf}\rightarrow\tau_\mathrm{sf}J_{-n}(\epsilon_\mathrm{ac}/\omega)$.
	On the other hand, if the particle tunnels from the left dot with spin down, it will absorb $n$ photons, so that $\tau_\mathrm{sf}\rightarrow\tau_\mathrm{sf}J_{n}(\epsilon_\mathrm{ac}/\omega)$.
	This situation is depicted schematically in \cref{fig:DQD_dark_state_chi_eps}(a).
	The effective Hamiltonian under an RWA then reads
	\begin{equation}
		\fl
		\hat{H}_\mathrm{RWA}^{(n)}=\left({\begin{array}{cccc}
				0 & 0 & -\tau_0J_0(\epsilon_\mathrm{ac}/\omega) & -\tau_{\mathrm{sf}}J_n(\epsilon_\mathrm{ac}/\omega) \\
				0 & 0 & \tau_{\mathrm{sf}}J_{-n}(\epsilon_\mathrm{ac}/\omega) & -\tau_0J_0(\epsilon_\mathrm{ac}/\omega)\\
				-\tau_0J_0(\epsilon_\mathrm{ac}/\omega) & \tau_{\mathrm{sf}}J_{-n}(\epsilon_\mathrm{ac}/\omega) & 0 & 0 \\
				-\tau_{\mathrm{sf}}J_{n}(\epsilon_\mathrm{ac}/\omega) & -\tau_0J_0(\epsilon_\mathrm{ac}/\omega) & 0 & 0
		\end{array}}\right).
		\label{eq:Hamiltonian_DS}
	\end{equation}
	When the number of photons is odd, the spin-flip term has the same sign for both spin directions because $J_{-n}(z) = (-1)^nJ_n(z)$.
	In the absence of magnetic field, time-reversal symmetry requires the spin-flip term to be proportional to the Pauli matrix $\hat{\tau}_y$, as discussed in \ref{sec:OME_term}.
	This form of tunneling prevents destructive interference between the spin-conserving and spin-flip tunneling channels.
	On the contrary, for odd numbers of emitted photons, the RWA Hamiltonian in \cref{eq:Hamiltonian_DS} has a spin-flip term of the form $\hat{\tau}_x\hat{\sigma}_x$, which does not exhibit this protection.
	As a result, a combination of nonzero $E_z$ (which breaks time-reversal symmetry) and nonzero $\epsilon_\mathrm{ac}$ (which leads to PAT) allows the spin-conserving and spin-flip paths to interfere destructively, leading to DSs.
	
	We search for DSs of the form $\ket{\mathrm{DS}}=\cos(\theta/2)\ket{L\uparrow}+e^{i\varphi}\sin(\theta/2)\ket{L\downarrow}$ with an arbitrary relative phase $\varphi$ and a mixing angle $\theta$.
	Using exact diagonalization of the Hamiltonian in \cref{eq:Hamiltonian_DS}, we find that DSs occur when the ratio $\chi$ between the spin-conserving and spin-flip amplitudes satisfies the following condition
	\begin{equation}
		\widetilde{\chi}^{(n)}=\frac{J_0(\epsilon_\mathrm{ac}/\omega)}{J_0(\epsilon_\mathrm{ac}/\omega) \pm i^{n+1}J_n(\epsilon_\mathrm{ac}/\omega)},
		\label{eq:condition_DS}
	\end{equation}
	where the superscript denotes the dependence of the DS on the resonance with the driving frequency $E_z=n\omega$.
	
	We impose $\widetilde{\chi}^{(n)}\in\mathds{R}$ (see \cref{eq:chi_SOC}), which means that the dark state occurs only when an odd number of photons are absorbed or emitted, i.e., $n = 2k + 1$ for $k\in \mathds{Z}$.
	In particular, for $E_z = 0$ ($n=0$), \cref{eq:condition_DS} does not have a real solution, highlighting the need to break time-reversal symmetry to observe DS formation.
	This dependence on the number of photons involved in the process is referred to as the odd-even effect and has been observed in DQDs, both theoretically and experimentally \cite{Burkard2002,Villavicencio2011,Stehlik2014,Zhou2021,Zhou2022}.
	This effect is typically intrinsic to multilevel systems, in which multiple pathways destructively interfere with each other \cite{Danon2014}.
	
	In \cref{fig:DQD_dark_state_chi_eps}(b-c) we show the current through a DQD in the cases of $n=1$ and $n=2$, respectively.
	In the former case, we plot the analytical prediction for the existence of a dark state, \cref{eq:condition_DS}, which is in agreement with the numerical results.
	If the spin-conserving tunneling path is not available, i.e., $\tau_0=0$, or equivalently $\chi=1$, then the only possible tunneling channel requires spin-flip, which is highly suppressed because of the large energy gap between states with opposite spin produced by $E_z$.
	This results in a current blockade unrelated to interference effects, and it is visible in the upper left corner of \cref{fig:DQD_dark_state_chi_eps}(b).
	For $n=2$, we observe that the only two situations where the current is suppressed coincide with $\chi\simeq 1$, $\epsilon_\mathrm{ac}=0$, and with $\chi=0$, $\epsilon_\mathrm{ac}\simeq 2.4\omega$, as expected from the even-odd effect.
	
	In an experimental device, in which the driving amplitude can be precisely tuned, the appearance of a sharp decrease in current at a given $\epsilon_\mathrm{ac}$ can be used to determine the value of $\chi$ and hence the SOC strength.	To minimize the uncertainty of the measurement, several odd resonances for $E_z=(2k + 1)\omega$ can be studied, providing a precise way to characterize the SOC present in a given device.
	
\begin{figure}
		\centering
		\includegraphics{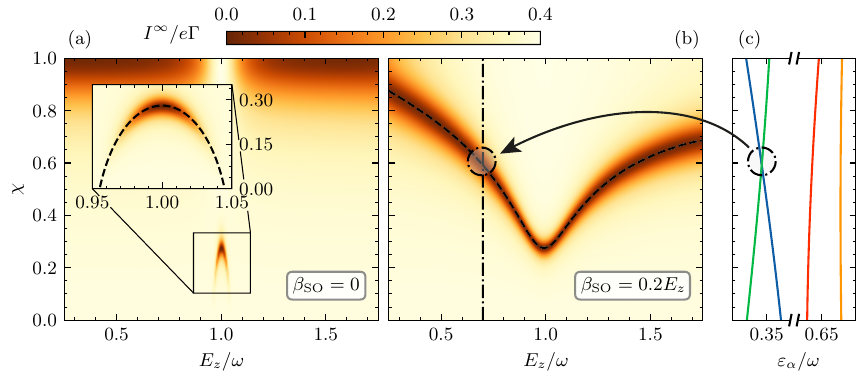}
		\caption{Current through a DQD as a function of $\chi$ and $E_z$ for (a) $\beta_\mathrm{SO} = 0$, and (b) $\beta_{\mathrm{SO}} = 0.2E_z$.
			Darker colors indicate the appearance of DSs.
			In panel (a), the inset shows a zoom into the DS near $E_z\simeq \omega$, and the dashed black line represents the analytical prediction given by equation~(\ref{eq:dark_state_approx}).
			In panel (b), the dashed black line represents the DS predicted by the crossing of Floquet quasienergies.
			Panel (c) shows the Floquet quasienergies for $E_z = 0.7\omega$ (dash-dotted line in panel (b)), with the energy crossing highlighted by a circle.
			The parameters used are $\delta=0$, $\omega = 10\tau = 100\Gamma$, and $\epsilon_{\mathrm{ac}}=2\omega$.}
		\label{fig:DQD_dark_state_Ez_chi}
	\end{figure}

	Next, we fix the driving amplitude at $\epsilon_\mathrm{ac}=2\omega$ and consider arbitrary values of $E_z$.
	The corresponding results are presented in \cref{fig:DQD_dark_state_Ez_chi}(a).
	The DS at the exact resonant condition $E_z = \omega$ extends to non-resonant values of $E_z$, forming a parabola in the $E_z-\chi$ plane.
	Again, we perform an exact diagonalization of the RWA Hamiltonian given by \cref{eq:Hamiltonian_DS} and look for dark states.
	Up to second order in $(E_z-\omega)$, these are found for the condition
	\begin{eqnarray}
		\widetilde{\chi}^{(1)} =& \frac{J_0(\epsilon_\mathrm{ac}/\omega)^2-|J_0(\epsilon_\mathrm{ac}/\omega)J_1(\epsilon_\mathrm{ac}/\omega)|}{J_0(\epsilon_\mathrm{ac}/\omega)^2 - J_1(\epsilon_\mathrm{ac}/\omega)^2} \nonumber\\
		&- \frac{(E_z-\omega)^2}{\tau^2 |J_0(\epsilon_\mathrm{ac}/\omega)J_1(\epsilon_\mathrm{ac}/\omega)|} + \mathcal{O}\left((E_z-\omega)^4\right).
		\label{eq:dark_state_approx}
	\end{eqnarray}
	The numerical results are consistent with our prediction, as shown in the inset of \cref{fig:DQD_dark_state_Ez_chi}(a).
	However, the validity of the RWA breaks down far from the resonance, rendering the equation above inapplicable.
	Nevertheless, even if the dot levels are in resonance at $\delta=0$, a trivial DS at $\chi=1$ emerges in which $E_z$ is no longer in resonance with the driving, leading to the particle being blocked in the left QD, as explained earlier.
	
	Moving on to the case of arbitrary OME term amplitudes, we observe a more complex situation.
	The OME term induces rapid spin rotations inside the QDs, which can prevent current blockade in scenarios where the particle would otherwise remain trapped in the left dot.
	This effect is particularly evident in \cref{fig:DQD_dark_state_Ez_chi} for $\chi\simeq 1$.
	When $\beta_\mathrm{SO}=0$ in \cref{fig:DQD_dark_state_Ez_chi}(a), the current is blocked if $E_z\neq\omega$, whereas if $\beta_\mathrm{SO}\neq0$ in \cref{fig:DQD_dark_state_Ez_chi}(b), a non-zero current can flow.
	However, despite the complex behavior, DSs can still be found, now extending to all values of $E_z$.
	Moreover, the steady-state spin projection is highly polarized, depending on whether $0<E_z<\omega$ (the particle remains in the left QD with spin down) or $\omega<E_z<2\omega$ (the final spin projection is inverted).
	Additionally, it is worth mentioning that the DSs obtained far from the resonance $E_z \neq \omega$ are highly pure, with $\Tr(\rho^2)\simeq 1$.
	These DSs could be used to store quantum information within a QD without requiring modification of the tunneling from the leads or between the dots.
	
	Since obtaining analytical results for the formation of DSs when $\beta_\mathrm{SO}\neq0$ in our system was not possible, we turned to the Floquet formalism for insight.
	For a Hamiltonian that is periodic with a period $T$, the wave function's time evolution can be expressed as $\ket{\Psi_\alpha(t)} = e^{-i\varepsilon_\alpha t}\ket{\phi_\alpha(t)}$, where $\varepsilon_\alpha$ are the quasienergies, and $\ket{\phi_\alpha}$ are the Floquet states obtained by diagonalizing the Floquet Hamiltonian $\hat{\mathcal{H}}(t) \equiv \hat{H}(t)-i\partial_t$.
	
	The Von Neumann-Wigner theorem \cite{Demkov2007} asserts that crossings between distinct energy levels occur only when the corresponding states belong to different representations of the system's symmetry group.
	Quasienergy crossings follow the same pattern and often indicate that the ac drive has restored the system's symmetry by suppressing relevant terms in the Hamiltonian~\cite{GomezLeon2011}.
	At these crossings, the time evolution operator depends exponentially on the difference between the quasienergies, and in resonance, the unitary time evolution matrix becomes the identity operator, freezing the long-term dynamics.
	
	Although the quasienergies for $\beta_\mathrm{SO}\neq0$ were obtained numerically \cite{Creffield2003}, their crossings match perfectly with the location of the DS found by evolving the Lindblad master equation (see \cref{fig:DQD_dark_state_Ez_chi}(b)).
	In \cref{fig:DQD_dark_state_Ez_chi}(c), we show the Floquet quasienergies for $E_z=0.7\omega$, where the crossing is at $\chi\simeq 0.6$, which agrees with the numerical result for the open system.
	
	\section{Flopping mode qubit operation}\label{sec:Out-of-res}
	We next focus on the situation in which the spin motion is confined to one of the dots and neglect the OME term for simplicity.
	In this regime, the spin dynamics is governed by the interplay between the Zeeman splitting and the TME as discussed in \cref{sec:theo-frame}.
	Our aim is to explore this regime as a possible implementation of a solid-state qubit.
	
	To obtain an effective Hamiltonian for an isolated spin in one of the dots with a constant electric field, we can perform a SWT.
	This transformation allows us to obtain the dynamics in one energy subspace that is weakly coupled and well separated in energy from another.
	In this case, we consider the states for each QD as each of the energy subspaces, so the effective Hamiltonian after the SWT is valid provided that we are far from the resonances (both spin-conserving and spin-flip ones), i.e., $|\delta|\gg\tau_0,|\delta\pm E_z|\gg\tau_\mathrm{sf}$.
	
	Under these conditions, we can obtain an effective Hamiltonian that is second order in the tunneling and contains all contributions due to virtual tunneling to the other dot.
	This effective Hamiltonian is given by
	\begin{eqnarray}
		\hat{H}_{\mathrm{eff}} &= \hat{H}_{\mathrm{eff}}^{(0)} + \hat{H}_{\mathrm{eff}}^{(2)},\\
		\hat{H}_{\mathrm{eff}}^{(0)} &= \frac{\delta}{2}\hat{\tau}_z + \frac{E_z}{2}\hat{\sigma}_z,\\
		\hat{H}_{\mathrm{eff}}^{\left(2\right)}&=\frac{\hat{\tau}_{z}}{2}(-\delta^{\left(2\right)}+b_{z}^{\left(2\right)}\hat{\sigma}_{z}+b_{\perp}^{\left(2\right)}\hat{\sigma}_{x}),
		\label{eq:effective_Hamiltonian_flopping}
	\end{eqnarray}
	where
	\numparts
	\begin{eqnarray}
		\delta^{\left(2\right)}&=\frac{2\tau_{0}^{2}}{\delta}+\tau_\mathrm{sf}^{2}\left(\frac{1}{\delta+E_z}+\frac{1}{\delta-E_{z}}\right),\\
		b_{z}^{\left(2\right)}&=\tau_\mathrm{sf}^2\left(\frac{1}{\delta+E_z}-\frac{1}{\delta-E_z}\right),\label{eq:Ez-out-of-reso}\\
		b_{\perp}^{\left(2\right)}&=\tau_{0}\tau_\mathrm{sf}\left(\frac{1}{\delta+E_z}-\frac{1}{\delta-E_{z}}\right).
		\label{eq:bx-out-of-reso}
	\end{eqnarray}
	\endnumparts
	The first term in equation~(\ref{eq:effective_Hamiltonian_flopping}) consists of renormalization of the detuning $\delta$ due to virtual tunneling.
	Similarly, the second term is a renormalization of $E_z$ which arises from spin-flip tunneling to the right dot and back.
	The third term arises due to the TME, which is an effective magnetic field perpendicular to the external field direction.
	This field arises from a combination of spin-conserving and spin-flip tunneling, where the spin virtually tunnels to the adjacent dot, flips its spin, and then tunnels back to the original dot through the spin-conserving path, resulting in an effective spin rotation.
	As discussed in \cref{sec:theo-frame}, the TME arises from the motion of the particle under the SOC (the inter-dot dynamics, in this case), and like the OME, requires $E_z\neq 0$ to break the time-reversal symmetry.
	
	The effective magnetic field induced by TME can be used in a similar manner to that of the flopping-mode qubit \cite{Croot2020,Teske2023}.
	However, the qubit cannot be manipulated via detuning alone due to the lack of two-axis control unless the spin-flip and tunneling amplitudes can be manipulated independently, i.e., unless $\chi$ itself can be tuned, e.g., by manipulating the overlap between the particle wave functions centered at each dot~\cite{Mutter2021}, rapidly enough to avoid dephasing.
	Nonetheless, for a time-dependent detuning, as discussed here, the TME allows for resonant manipulation of the qubit.
	
\begin{figure}
		\centering
		\includegraphics{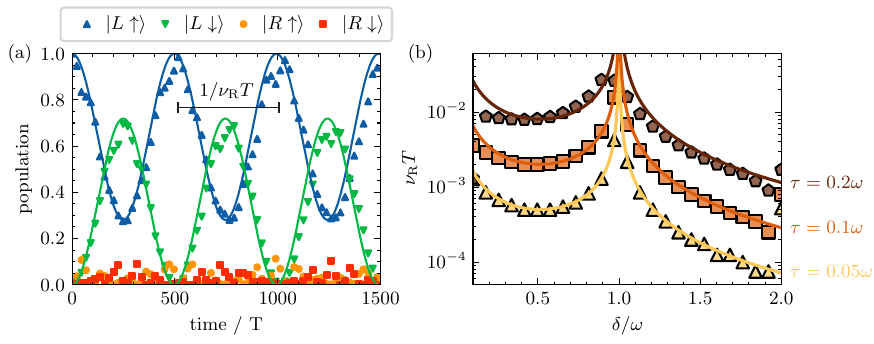}
		\caption{Panel (a) shows the Rabi oscillations in a closed system described by the full Hamiltonian given by \cref{eq:total_hamiltonian}, plotted as symbols.
			For comparison, the dynamics given by the RWA Hamiltonian (see \cref{eq:effective_Hamiltonian_RWA}) is also shown as solid lines.
			The Rabi frequency is denoted by $\nu_\mathrm{R}$ and is shown in the figure.
			The parameters used in this panel are $\delta=\omega/2$ and $\tau=0.1\omega$.
			In panel (b), the Rabi frequency is shown as a function of detuning for different total tunneling rates, with results obtained using both the full Hamiltonian (symbols) and the effective RWA Hamiltonian (solid lines).
			The total tunneling rate is $\tau=0.05\omega$ (yellow, triangles), $\tau=0.1\omega$ (orange, squares), and $\tau=0.2\omega$ (brown, pentagons).
			Other parameters, common for both panels, are $\epsilon_\mathrm{ac} = 0.2\omega$, $E_z=\omega$, $\chi=0.2$, and $\beta_\mathrm{SO}=0$.}
		\label{fig:flopping_mode_energies}
	\end{figure}

	Let us consider the case where the amplitude of the ac gate signal is much smaller than the driving frequency, i.e., $\epsilon_\mathrm{ac}\ll\omega$.
	In this regime, we can apply the RWA and obtain an effective spin model for the resonance condition $ E_z=\omega$.
	By applying the ac gate with a phase $\phi$, we can achieve two-axis control of the qubit, allowing for coherent manipulation of the qubit state
	\begin{equation}
		\hat{H}_{\mathrm{RWA}}=\frac{b_{z}^{\left(2\right)}}{2}\hat{\sigma}_z-\frac{\tilde{b}_{1,\perp}}{2}\left(\cos\phi\,\hat{\sigma}_x+\sin\phi\,\hat{\sigma}_y\right),
		\label{eq:effective_Hamiltonian_RWA}
	\end{equation}
	where
	\begin{equation}
		\tilde{b}_{1,\perp}=\frac{4\epsilon_{\mathrm{ac}}\tau_{0}\tau_\mathrm{sf}E_{z}}{\delta\left(\delta^{2}-E_{z}^{2}\right)}.
		\label{eq:RWA-amp}
	\end{equation}
	Otherwise, the frequency of the ac gate can be matched with the renormalized splitting of the two levels of the qubit $E_z + b_z^{(2)}$.
	The correction to $\tilde{b}_{1,\perp}$ in this case can be given as 
	\begin{equation}
		\tilde{b}_{1,\perp}\to
		\tilde{b}_{1,\perp} + \frac{2\epsilon_{\mathrm{ac}}\tau_{0}\tau_\mathrm{sf}b_{z}^{\left(2\right)}}{\left(\delta^{2}-E_{z}^{2}\right)^{2}},
	\end{equation}
	which is small in the context of the previous approximations.
	In \cref{fig:flopping_mode_energies}(a), we can see the dynamics of the flopping-mode qubit using the effective Hamiltonian described above.
	For comparison, we also show the results obtained by numerically integrating the equation of motion of a closed system under the original Hamiltonian given by \cref{eq:total_hamiltonian}.
	Since both dot energy levels are out of resonance, with $\delta=\omega / 2$, the population of the right dot is small and the particle remains in the left dot.
	The results obtained demonstrate that the effective model accurately reproduces the dynamics of the system.
	
	The Rabi oscillation frequency of the effective model is given by
	\begin{equation}
		\nu_\mathrm{R} = \frac{1}{2\pi}\sqrt{\left(b_z^{(2)}\right)^2+\left(\tilde{b}_{1,\perp}\right)^2}.
	\end{equation}
	We compare the Rabi frequency of the effective RWA Hamiltonian with that obtained from the original Hamiltonian; see \cref{fig:flopping_mode_energies}(b).
	As expected, the results agree well for $|\delta - n\omega|\gg\tau$.
	
	However, this method of manipulation couples the spin with electric fluctuations due to the dependence of $b_z^{(2)}$ and $b_\perp^{(2)}$ on $\delta$.
	In purified silicon, where the magnetic noise caused by the atomic nuclei is heavily suppressed, this may be the main source of decoherence for the qubit~\cite{Tyryshkin2011,Veldhorst2014}.
	Unfortunately, in the low ac gate amplitude described here, the flopping-mode qubit lacks a natural sweetspot~\cite{Makhlin2004,Fei2015} where the system is insensitive to electric noise to first order in the coupling to the bath.
	In particular, electric noise that enters through $b_z^{(2)}$ could only be efficiently suppressed at $\delta=0$, where the first derivative with respect to $\delta$, i.e., the noise susceptibility, vanishes.
	However, at this value, the qubit cannot be operated as the direct spin-conserving transition is resonant.
	Fortunately, the perpendicular component of the effective magnetic field of \cref{eq:RWA-amp} can be made insensitive to noise to first order at $\delta = \pm E_z/\sqrt{3}$, which can improve qubit operation.
	This dynamical sweetspot~\cite{PicoCortes2019,PicoCortes2021} is observed as a minimum of the Rabi frequency at $\delta \simeq 0.58\omega$ in \cref{fig:flopping_mode_energies}(b).
	
	In \ref{sec:TME-arbitrary-amps}, we provide expressions for the TME at arbitrary ac detuning amplitudes.
	Unlike the simple limit shown in \cref{eq:RWA-amp}, these expressions are quite complex, highlighting the fact that the effective magnetic field arises from second-order virtual tunneling involving two PAT processes.
	In the arbitrary-amplitude limit, as given in \ref{sec:TME-arbitrary-amps}, we can achieve a fine degree of spin manipulation.
	Furthermore, we can also obtain dynamical sweetspots for large ac gate amplitudes.
	
	\section{Conclusions}\label{sec:conclusions}
	We investigated the influence of spin-orbit coupling (SOC) on spin transport in a periodically driven double quantum dot (DQD) system.
	Due to the presence of a spin-flip tunneling path, the onsite energy difference between the dots can be tuned to achieve highly spin-polarized currents.
	In this work, we propose several mechanisms for obtaining highly polarized currents in both directions with fully electric control, enabling fast switching of the polarization under experimental conditions.
	
	The combination of the orbital dynamics inside the dots and the applied ac voltage results in the appearance of an effective magnetic field that drives electric dipole spin resonances (EDSR) in the dots.
	The effect of this field can be most clearly observed in the appearance of characteristic avoided crossings of the tunneling resonances near the EDSR condition, where the ac voltage frequency is approximately equal to the Zeeman splitting.
	Under this condition, tunneling results in novel phenomena, such as spin-dependent tunneling amplitudes due to the constructive self-interference of the sidebands originating from the electric and effective magnetic fields.
	Remarkably, both of these processes arise from the same ac voltage, with the electric field resulting directly from the voltage, and the effective magnetic field incorporating the dynamics of the excited states of the potential.
	
	We also investigate the appearance of new dark states when the system is driven under the EDSR condition, resulting from the interference between photo-assisted spin-conserving and spin-flip processes with different numbers of photons involved.
	In the absence of a Zeeman splitting, time-reversal symmetry prevents destructive interference between these two processes.
	However, in the presence of a magnetic field, together with an ac voltage, these two paths can interfere and create dark states.
	These states exhibit a characteristic even-odd effect, appearing only for odd sideband transitions.
	The current drop at the location of these dark states is very sharp, allowing for the characterization of the SOC present in the system.
	These dark states could be valuable for quantum information storage due to their high purity and spin polarization.
	
	Finally, we investigate the viability of flopping-mode qubit operations when the particle is localized in one of the dots.
	We provide expressions for the effective magnetic field arising from the interdot motion of the particle under SOC for arbitrary ac-gate amplitudes and focus on the experimentally accessible case of small gate amplitudes.
	In this situation, we identify a dynamical sweetspot induced by the ac voltage, where the Rabi frequency is insensitive to charge noise to first order in detuning.
	This point of operation may be significant for novel solid-state quantum computing platforms, such as isotopically purified silicon, where decoherence due to the nuclear magnetic field can be suppressed, and electric noise may be the most relevant source of decoherence.
	
	\ack
	G.P. and D.F.F. are supported by Spain’s MINECO through Grant No. PID2020-117787GB-I00 and by the CSIC Research Platform PTI-001.
    G.P. and D.F.F. also acknowledge the agreement between Carlos III University and the CSIC through the UA.
	D.F.F. acknowledges support from FPU Program No. FPU20/04762.
	J.P.C. acknowledges DFG funding through project B04 of SFB 1277 Emerging Relativistic Phenomena in Condensed Matter.
	D.F.F. and J.P.C. contributed equally to this work.
	
	\appendix
	
	\section{Effective model}\label{sec:OME_term}
	In this appendix, we provide a brief overview of the origin of the OME term $\hat{H}_1(t)$ in the Hamiltonian of equation~(\ref{eq:H_total}).
	To obtain the effective Hamiltonian on the basis employed in the main text, we follow a similar derivation to \cite{Rashba2008,Borhani2012} and employ a Schrieffer-Wolff transformation (SWT).
	We start by considering the Hamiltonian for a single particle in a linear DQD (in the $x$-direction) under both electric and magnetic fields and in the presence of SOC modeled by the spin-orbit vector $\bm{\alpha}=(\alpha_x,\alpha_y,0)$~\cite{Villavicencio2013}, that is compatible with a two-dimensional electron gas grown along the [001] direction~\cite{Hanson2007}.
	In our model, the $z$-direction corresponds to the direction normal to the QD plane.
	We consider an electric field in the $x$-direction and a Zeeman splitting $E_{z}$ in the $z$-direction.
	The resulting Hamiltonian can be written as
	\begin{eqnarray}
		\hat{H}\left(x,t\right) =\hat{H}_{k}+V\left(x\right)+\hat{H}_{e}\left(x,t\right)+\hat{H}_{z}+\hat{H}_{\mathrm{SO}},\\
		\hat{H}_{k} =k^{2}/2m,\\
		\hat{H}_{e}\left(x,t\right)=exE\left(t\right),\\
		\hat{H}_{z} =E_{z}\hat{\sigma}_{z}/2,\\
		\hat{H}_{\mathrm{SO}} =\bm{\alpha}\cdot\hat{\bm{\sigma}}k.
	\end{eqnarray}
	The scalar potential $V\left(x\right)$ for the DQD exhibits two minima at $\pm\ell$, near which the potential can be chosen as harmonic $V_{\mathrm{osc}}\left(x\right)=\left(1/2\right)m\omega_{0}^{2}x^{2}$.
	In the tight-binding approximation, we first evaluate the local Hamiltonians in each dot in the eigenfunctions of the individual harmonic potentials
	\begin{equation}
		\ket{\psi_{\eta,\nu,\sigma}}=\ket{\eta,\nu}\ket{\sigma},
	\end{equation}
	with $\nu\in\mathbb{N}$ labeling the eigenstates of the harmonic potential and $\sigma\in\left\{\uparrow,\downarrow\right\}$ the spin projection along the z-axis.
	The orbital part $\ket{\eta,\nu}$ of these eigenfunctions can be obtained by diagonalizing the Hamiltonian $\hat{H}_{k}+V_{\mathrm{osc}}\left(x-\eta\ell\right)$, with $\eta=\pm$ corresponding to the left and right dots, as appropriate, and corresponds to a Fock-Darwin function with shifted centers.
	Following this, around $\eta\ell$ we can write the terms of the Hamiltonian as
	\begin{eqnarray}
		\hat{H}_{k}+V_{\mathrm{osc}}\left(x-\eta\ell\right) =\omega_{0}\left(\hat{a}^{\dagger}\hat{a}+1\right),\\
		\hat{H}_{e}\left(\eta\ell,t\right) =el_{0}E\left(t\right)\left(\hat{a}^{\dagger}+\hat{a}\right)+e\eta E\left(t\right)\ell,\\
		\hat{H}_{\mathrm{SO}} =\frac{i}{2l_{0}}\bm{\alpha}\cdot\hat{\bm{\sigma}}\left(\hat{a}^{\dagger}-\hat{a}\right),
	\end{eqnarray}
	with $\hat{a}^{\dagger},\hat{a}$ the Fock operators of the oscillator and the Zeeman term unchanged.
	Moreover, we employ the characteristic oscillator length $l_{0}=\sqrt{\omega_{0}/2m}$.
	We can separate this Hamiltonian into a part that is static in the orbital dynamics
	\begin{equation}
		\hat{H}_{\eta}^{\left(0\right)}=\omega_{0}\left(\hat{a}^{\dagger}\hat{a}+1\right)+eE\left(t\right)\eta\ell+\left(E_{z}/2\right)\hat{\sigma}_{z},
	\end{equation}
	and a dynamic part
	\begin{eqnarray}
		\hat{H}_{\eta}^{\left(1\right)} & =el_{0}E\left(t\right)\left(\hat{a}^{\dagger}+\hat{a}\right)+(i/2l_{0})\bm{\alpha}\cdot\hat{\bm{\sigma}}\left(\hat{a}^{\dagger}-\hat{a}\right).
	\end{eqnarray}
	When projected into the ground state, the first part reduces to the usual description of single-state QDs.
	However, if we project the second part as well, we will obtain an effective Hamiltonian that incorporates the action of the SOC into the ground state.
	We do this by first performing a SWT $\hat{H}_{\eta}^{\prime}\left(t\right)=e^{\hat{\Upsilon}}\hat{H}_{\eta}\left(t\right)e^{-\hat{\Upsilon}}$.
	We further consider the adiabatic approximation with respect to the electric field, valid provided that the driving frequency is much lower than the oscillator frequency, i.e., $\omega\ll\omega_0$.
	Then the anti-Hermitian operator $\hat{\Upsilon}$ is given by
	\begin{equation}
		\left[\hat{H}_{\eta}^{\left(0\right)},\hat{\Upsilon}\right]=\hat{H}_{\eta}^{\left(1\right)},
	\end{equation}
	resulting in
	\begin{equation}
		\hat{\Upsilon}=\left(f+\bm{d}\cdot\hat{\bm{\sigma}}\right)\hat{a}^{\dagger}-\left(f^{*}+\bm{d}^{*}\cdot\hat{\bm{\sigma}}\right)\hat{a},
	\end{equation}
	where
	\begin{eqnarray}
		f & =el_{0}E_{\eta}\left(t\right)/\omega_{0},\\
		d_{x} & =\frac{1}{2l_{0}}\frac{i\omega_{0}\alpha_{x}+\alpha_{y}E_{z}}{\omega_{0}^{2}-E_{z}^{2}},\\
		d_{y} & =\frac{1}{2l_{0}}\frac{i\omega_{0}\alpha_{y}-\alpha_{x}E_{z}}{\omega_{0}^{2}-E_{z}^{2}}, \\
		d_{z} &=0.
	\end{eqnarray}
	The effective action of the SOC term in the ground state is described by the Hamiltonian
	\begin{eqnarray}
		\fl
		\hat{H}_{\eta}^{\left(2\right)}=\frac{1}{2}\left[\hat{\Upsilon},\hat{H}_{\eta}^{\left(1\right)}\right]
		=\frac{1}{2} &\left\{ \left[ \left(f+\bm{d}\cdot\hat{\bm{\sigma}}\right),\left(el_{0}E\left(t\right)+(i/2l_{0})\bm{\alpha}\cdot\hat{\bm{\sigma}}\right)\right]\hat{a}^{\dagger}\hat{a}^{\dagger}\right. \nonumber \\
		&-\left[\left(f^{*}+\bm{d}^{*}\cdot\hat{\bm{\sigma}}\right),\left(el_{0}E\left(t\right)-(i/2l_{0})\bm{\alpha}\cdot\hat{\bm{\sigma}}\right)\right]\hat{a}\hat{a} \nonumber \\
		&+ \left[\left(f+\bm{d}\cdot\hat{\bm{\sigma}}\right),\left(el_{0}E\left(t\right)-(i/2l_{0})\bm{\alpha}\cdot\hat{\bm{\sigma}}\right)\right]\hat{a}^{\dagger}\hat{a} \nonumber \\
		& +\left(el_{0}E\left(t\right)-(i/2l_{0})\bm{\alpha}\cdot\hat{\bm{\sigma}}\right)\left(f+\bm{d}\cdot\hat{\bm{\sigma}}\right) \nonumber \\
		&+ \left[\left(el_{0}E\left(t\right)+(i/2l_{0})\bm{\alpha}\cdot\hat{\bm{\sigma}}\right),\left(f^{*}+\bm{d}^{*}\cdot\hat{\bm{\sigma}}\right)\right]\hat{a}^{\dagger}\hat{a} \nonumber \\
		&\left. +\left(f^{*}+\bm{d}^{*}\cdot\hat{\bm{\sigma}}\right)\left(el_{0}E\left(t\right)+(i/2l_{0})\bm{\alpha}\cdot\hat{\bm{\sigma}}\right)\right\}.
	\end{eqnarray}
	The two-pair excitation processes $\left(\propto \hat{a}\hat{a},\hat{a}^{\dagger}\hat{a}^{\dagger}\right)$ connect states that are separated in energy by $2\omega_{0}$ and therefore can be neglected in our effective Hamiltonian approximation, while the terms $\propto \hat{a}^{\dagger}\hat{a}$ do not contribute to the energy of the ground state.
	Hence, we find the following
	\begin{equation}
		\hat{H}_{\eta}^{\left(2\right)}=
		\frac{-E_{z}||\bm{\alpha}||^2}{2l_{0}^{2}\left(\omega_{0}^{2}-E_{z}^{2}\right)}\hat{\sigma}_{z}
		+\frac{E_{z}eE(t)}{\omega_{0}^{2}-E_{z}^{2}}\bm{\alpha}^\perp\cdot\hat{\bm{\sigma}}.
		\label{eq:Heta2}
	\end{equation}
	where $\bm{\alpha}^\perp=(\alpha_y,-\alpha_x,0$).
	The first term shifts the Zeeman splitting to 
	\begin{equation}
		\widetilde{E}_z = E_z\left(1 - \frac{||\bm{\alpha}||^2}{2l_{0}^{2}\left(\omega_{0}^{2}-E_{z}^{2}\right)}\right),
	\end{equation}
	while the second term is the OME field that we sought.
	Crucially, it is oriented perpendicular to the direction of the SOC field, i.e., to $\bm{\alpha}$.
	Note that both terms require $E_{z}\neq0$ to break the time-reversal symmetry.
	Moreover, the constant part of the electric field $E(t)$ in the second term will rotate the spin quantization axis.
	However, this rotation is produced around the direction determined by $\bm{\alpha}$.
	As shown in the following, the spin-flip amplitude is aligned in this direction and is unaffected by this rotation.
	The Zeeman splitting along the new quantization axis is given by 
	\begin{equation}
		\widetilde{E}_z\to \widetilde{E}_z\sqrt{1+\frac{e^2E_0^2||\bm{\alpha}||^4}{(\omega_0^2-E_z^2)^2}}\approx \widetilde{E}_z\left(1+\frac{e^2E_0^2||\bm{\alpha}||^4}{2(\omega_0^2-E_z^2)^2}\right),
	\end{equation}
	with $E_0$ being the constant part of the electric field.
	This is a next-order effect compared to the shift from $E_z$ to $\widetilde{E}_z$ and can be ignored.
	The OME term is similarly rotated, but this is also a higher-order effect, and we disregard it as well.
	
	Regarding tunneling amplitudes, we consider orthonormal Wannier functions of the ground state of each dot, defined as~\cite{Mutter2021}
	\begin{equation}
		\ket{w_{\eta,\sigma}} =\frac{1}{\sqrt{N}}(\ket{\psi_{\eta,0,\sigma}} + \gamma\ket{\psi_{\bar{\eta},0,\sigma}}),
	\end{equation}
	where $N \equiv 1 - 2\gamma S + \gamma^2$, $\gamma\equiv(1-\sqrt{1-S^2})/S$, and $S \equiv \braket{\psi_{L,0,\sigma}}{\psi_{R,0,\sigma}}$ is the overlap between the dot wave functions.
	Regardless of the particularities of $V(x)$, we can consider a standard real-valued tunneling matrix element $\tau_0$ without loss of generality.
	The spin-flip tunneling amplitude can be obtained as 
	
	\begin{equation}
		\tau_{\mathrm{sf}} = \bra{w_{L\sigma}}\hat{H}_\mathrm{SO}\ket{w_{R\sigma'}} 
		= \frac{1-\gamma^2}{N}\bra{\sigma}\bm{\alpha}\cdot\hat{\bm{\sigma}}\ket{\sigma'}\bra{L,0}k\ket{R,0},
	\end{equation}
	with $\ket{\eta,0}$ the ground state of the respective harmonic oscillator, as defined above.
	The general form of the expected value $\bra{L,0}k\ket{R,0}$ can be determined by imposing time-reversal invariance of the spin-orbit Hamiltonian, i.e., $\mathcal{T}\hat{H}_\mathrm{SO}\mathcal{T}^{-1} = \hat{H}_\mathrm{SO}$.
	In the most general way, the SOC Hamiltonian, written on the basis of $\left\{\ket{L\uparrow}, \ket{L\downarrow},\ket{R\uparrow}, \ket{R\downarrow}\right\}$, reads as follows
	\begin{equation}
		\hat{H}_\mathrm{SO}=\left({\begin{array}{cccc}
				0 & 0 & 0 & \tau_{\mathrm{sf}}\\
				0 & 0 & -\tau_{\mathrm{sf}}^* & 0\\
				0 & -\tau_{\mathrm{sf}} & 0 & 0\\
				\tau_{\mathrm{sf}}^* & 0 & 0 & 0
		\end{array}}\right).
	\end{equation}
	Taking $\bm{\alpha}$ along the $y$-direction, as considered throughout this work, we recover a term $\propto\hat{\tau}_y\hat{\sigma}_y$, which yields a real spin-flip matrix of the form
	\begin{equation}
		\hat{H}_\mathrm{SO}=\left({\begin{array}{cccc}
				0 & 0 & 0 & \tau_{\mathrm{sf}}\\
				0 & 0 & -\tau_{\mathrm{sf}} & 0\\
				0 & -\tau_{\mathrm{sf}} & 0 & 0\\
				\tau_{\mathrm{sf}} & 0 & 0 & 0
		\end{array}}\right).
	\end{equation}
	The fact that the spin-flip tunneling term is $\propto\hat{\tau}_y$ is crucial.
	Otherwise, the spin-flip and spin-conserving tunneling terms can interfere destructively.
	Consider, for instance, that the tunneling is of the form $\propto\hat{\tau}_x\hat{\sigma}_x$.
	By performing a $\pi/2$ rotation around the y-axis, we obtain
	\begin{equation}
		\hat{H}_\mathrm{T}=\left({\begin{array}{cccc}
				0 & 0 & -\tau_0 + \tau_\mathrm{sf} & 0\\
				0 & 0 & 0 & -\tau_0 - \tau_\mathrm{sf}\\
				-\tau_0 + \tau_\mathrm{sf} & 0 & 0 & 0\\
				0 & -\tau_0 - \tau_\mathrm{sf} & 0 & 0
		\end{array}}\right),
	\end{equation}
	which can exhibit a dark state when $\tau_0 = \tau_\mathrm{sf}$, i.e., when $\chi = 0.5$.
	However, if, as here, we have a term $\propto\hat{\tau}_y\hat{\sigma}_y$, the rotation yields
	\begin{equation}
		\hat{H}_\mathrm{T}=\left({\begin{array}{cccc}
				0 & 0 & -\tau_0 - i\tau_\mathrm{sf} & 0\\
				0 & 0 & 0 & -\tau_0 +i\tau_\mathrm{sf}\\
				-\tau_0 + i\tau_\mathrm{sf} & 0 & 0 & 0\\
				0 & -\tau_0 - i\tau_\mathrm{sf} & 0 & 0
		\end{array}}\right),
	\end{equation}
	preventing destructive interference.
	
	\section{TME for arbitrary ac amplitudes}\label{sec:TME-arbitrary-amps}
	In this appendix, we give the expressions for the TME terms in the time-dependent case with arbitrary ac amplitudes.
	After a time-dependent SWT~\cite{Goldin2000}, we obtain an effective Hamiltonian up to second order in the tunnel couplings, given by
	\begin{equation}
		\hat{H}_{\mathrm{eff}}^{\left(2\right)}\left(t\right)=
		\frac{\hat{\tau}_{z}}{2}\{-\delta^{\left(2\right)}\left(t\right)+b_{z}^{\left(2\right)}\left(t\right)\hat{\sigma}_{z}+[ b_{x}^{\left(2\right)}\left(t\right)\hat{\sigma}_{x}+b_{y}^{\left(2\right)}\left(t\right)\hat{\sigma}_{y}]\},
	\end{equation}
	where the time-dependent detuning and Zeeman splittings are given by
	\begin{eqnarray}
		\delta^{\left(2\right)}\left(t\right)&=\sum_{\mu,\nu}J_{\mu}\left(\frac{\epsilon_{\mathrm{ac}}}{\omega}\right)J_{\nu}\left(\frac{\epsilon_{\mathrm{ac}}}{\omega}\right)\cos[(\mu-\nu)\omega t]\nonumber\\
		&\times\left[\frac{2\tau_{0}^{2}}{\delta-\nu\omega}+\frac{\tau_\mathrm{sf}^{2}}{\delta+E_{z}+\nu\omega}+\frac{\tau_\mathrm{sf}^{2}}{\delta-E_{z}+\nu\omega}\right],\\
		b_{z}^{\left(2\right)}\left(t\right)&=\sum_{\mu,\nu}J_{\mu}\left(\frac{\epsilon_{\mathrm{ac}}}{\omega}\right)J_{\nu}\left(\frac{\epsilon_{\mathrm{ac}}}{\omega}\right)\cos[(\mu-\nu)\omega t]\nonumber\\
		&\times\left[\frac{\tau_\mathrm{sf}^2}{\delta+E_{z}+\nu\omega}-\frac{\tau_\mathrm{sf}^2}{\delta-E_{z}+\nu\omega}\right].
	\end{eqnarray}
	The magnetic field gradient in the perpendicular direction is given by
	\begin{eqnarray}
		b_{x}^{\left(2\right)}\left(t\right)&=\sum_{\mu,\nu}J_{\mu}\left(\frac{\epsilon_{\mathrm{ac}}}{\omega}\right)J_{\nu}\left(\frac{\epsilon_{\mathrm{ac}}}{\omega}\right)\cos[(\mu-\nu)(\omega t+\phi)]\nonumber\\&\times\left[\frac{\tau_{0}\tau_\mathrm{sf}}{\delta+E_{z}-\nu\omega}-\frac{\tau_{0}\tau_\mathrm{sf}}{\delta-E_{z}-\nu\omega}\right]\\
		b_{y}^{\left(2\right)}\left(t\right)&=\sum_{\mu,\nu}J_{\mu}\left(\frac{\epsilon_{\mathrm{ac}}}{\omega}\right)J_{\nu}\left(\frac{\epsilon_{\mathrm{ac}}}{\omega}\right)\sin[(\mu-\nu)(\omega t+\phi)]\nonumber\\&\times\left(\frac{2\tau_{0}\tau_\mathrm{sf}}{\delta-\nu\omega}-\frac{\tau_{0}\tau_\mathrm{sf}}{\delta+E_{z}-\nu\omega}-\frac{\tau_{0}\tau_\mathrm{sf}}{\delta-E_{z}-\nu\omega}\right).
	\end{eqnarray}
	Note that these expressions involve two different photon numbers $\mu$ and $\nu$, as they are virtual second-order tunneling processes that involve two photo-assisted transitions~\cite{GallegoMarcos2015,PicoCortes2019}.
	
	In an $n-$photon resonance $E_z\simeq n\omega$, the Hamiltonian in the RWA is given by
	\begin{equation}
		\hat{H}_{n}^{\left(2\right)}\left(t\right)=\frac{\hat{\tau}_{z}}{2}\{-\widetilde{\delta}^{\left(2\right)} +(E_z + \widetilde{b}_{z}^{\left(2\right)}- n\omega)\hat{\sigma}_{z}+\widetilde{b}_{n,\perp}^{\left(2\right)}[\cos(n\phi)\hat{\sigma}_{x}+\sin(n\phi)\hat{\sigma}_{y}]\},
		\label{eq:RWA_Hamiltonin_flopping_mode}
	\end{equation}
	where the diagonal terms are given by
	\begin{eqnarray}
		\widetilde{\delta}^{\left(2\right)}&=\sum_{\nu}J_{\nu}^{2}\left(\frac{\epsilon_{\mathrm{ac}}}{\omega}\right)\left[\frac{2\tau_{0}^{2}}{\delta-\nu\omega}+\frac{\tau_\mathrm{sf}^{2}}{\delta+E_{z}+\nu\omega}+\frac{\tau_\mathrm{sf}^{2}}{\delta-E_{z}+\nu\omega}\right],\\
		\widetilde{b}_{z}^{\left(2\right)}&=\sum_{\nu}J_{\nu}^{2}\left(\frac{\epsilon_{\mathrm{ac}}}{\omega}\right)\left(\frac{\tau_\mathrm{sf}^2}{\delta+E_{z}+\nu\omega}-\frac{\tau_\mathrm{sf}^2}{\delta-E_{z}+\nu\omega}\right),
	\end{eqnarray}
	and the off-diagonal term has an amplitude
	\begin{eqnarray}
		\widetilde{b}_{n,\perp}^{(2)}&=\sum_{\nu}J_{\nu}\left(\frac{\epsilon_{\mathrm{ac}}}{\omega}\right)J_{\nu+n}\left(\frac{\epsilon_{\mathrm{ac}}}{\omega}\right)
		\left(\frac{\tau_{0}\tau_\mathrm{sf}}{\delta-\nu\omega}-\frac{\tau_{0}\tau_\mathrm{sf}}{\delta-E_{z}-\nu\omega}\right)
		\nonumber\\&
		-\sum_{\nu}J_{\nu}\left(\frac{\epsilon_{\mathrm{ac}}}{\omega}\right)J_{\nu-n}\left(\frac{\epsilon_{\mathrm{ac}}}{\omega}\right)\left(\frac{\tau_{0}\tau_\mathrm{sf}}{\delta-\nu\omega}-\frac{\tau_{0}\tau_\mathrm{sf}}{\delta+E_{z}-\nu\omega}\right).
	\end{eqnarray}
	
	\begin{figure}
		\centering
		\includegraphics{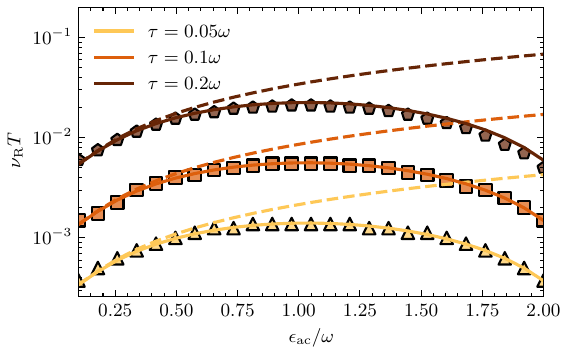}
		\caption{Rabi frequency as a function of driving amplitude for different total tunneling rates.
			The results are obtained using three different Hamiltonians: the full Hamiltonian (represented by symbols), the low-amplitude effective RWA Hamiltonian given by \cref{eq:effective_Hamiltonian_RWA} in the main text (represented by dashed lines), and the RWA Hamiltonian taking into account up to five terms in the infinite series given by \cref{eq:RWA_Hamiltonin_flopping_mode} (represented by solid lines).
			The total tunneling rates considered are $\tau=0.05\omega$ (yellow, triangles), $\tau=0.1\omega$ (orange, squares), and $\tau=0.2\omega$ (brown, pentagons).
			The other parameters used are $E_z=\omega$, $\delta=\omega/2$, $\chi=0.2$, and $\beta_\mathrm{SO}=0$.}
		\label{fig:higher_RWA}
	\end{figure}
	
	In \cref{fig:higher_RWA}, we compare the Rabi frequency of a flopping-mode qubit obtained using the full Hamiltonian with the results given by the effective RWA Hamiltonian shown above.
	To obtain numerical results, we truncated the summations to $\nu=-2, -1, \ldots, 2$.
	Remarkably, we find that both results agree well, even when working with tunneling rates as high as $\tau=0.2\omega$.
	We also compare our results with the prediction given by the low-amplitude effective Hamiltonian shown in \cref{eq:effective_Hamiltonian_RWA} of the main text.
	In the limit of $\epsilon_\mathrm{ac}<\omega/2$, all the results coincide.
	
	\section{TME under the spin resonance condition: asymmetry in \texorpdfstring{$\delta$}{δ}}\label{sec:TME-spin-resonance}
	To gain insight into the asymmetry in $\delta$ discussed in \cref{sec:DQD_global_OME} of the main text, we study the TME under spin resonance.
	We consider the Hamiltonian of \cref{eq:HRWA_2} in a situation where direct tunneling between the two dots is energetically disfavored, which can be achieved by lowering $\tau$.
	Since in this appendix we are concerned with a qualitative understanding of the asymmetry, we do not regard the precise conditions of validity and always assume a value of $\tau$ that makes the SWT valid.
	Under this condition, virtual tunneling processes are dominant, and we can employ a SWT to obtain an effective Hamiltonian with the leading term being of second order in the tunneling amplitudes, as done above for the flopping-mode operation in \cref{sec:Out-of-res} of the main text.
	We apply these transformations in the frame discussed in \cref{sec:DQD_global_OME} leading to \cref{eq:HRWA_2}, i.e., after applying \cref{eq:Ut-prime} and \cref{eq:Ut}) and working in the RWA.
	This is different from the treatment of \ref{sec:TME-arbitrary-amps}, where SWT was applied before RWA, which amounts to neglecting photo-assisted virtual tunneling processes.
	However, a complete treatment of the TME in the presence of the OME lies outside the scope of this work.
	
	In this rotating frame, as discussed above, the $\beta_\mathrm{SO}$ term plays a role analogous to the Zeeman splitting.
	Applying the transformation yields the following effective spin model for the left dot
	\begin{equation}
		\hat{H}_{\mathrm{eff}} = \frac{1}{2}\left(\widetilde{E}_z + \widetilde{b}_{z}^{\left(2\right)}\right)\hat{\sigma}_{z}+\frac{\tilde{b}_{\perp}^{(2)}}{2}\hat{\sigma}_{x},
		\label{eq:effective_Hamiltonian-TME-SR}
	\end{equation}
	where the effective magnetic field is given by 
	\begin{eqnarray}
		\tilde{b}_z^{(2)} = &\frac{\widetilde{\tau}_\downarrow^2-\widetilde{\tau}_\uparrow^2}{\delta} +\frac{2\widetilde{E}_z\widetilde{\tau}_\mathrm{sf}^2}{\widetilde{E}_z^2-\delta^2},\\ 
		\tilde{b}_\perp^{(2)} = &\widetilde{\tau}_\mathrm{sf}\left(\frac{\widetilde{\tau}_\downarrow}{\widetilde{E}_z-\delta}-\frac{\widetilde{\tau}_\uparrow}{\widetilde{E}_z+\delta}-\frac{\widetilde{\tau}_\uparrow+\widetilde{\tau}_\downarrow}{\delta}\right).
	\end{eqnarray}
	The term $\widetilde{b}_z^{(2)}$ is non-zero only when there are spin-dependent tunneling amplitudes $\abs{\widetilde{\tau}_\downarrow} \neq \abs{\widetilde{\tau}_\downarrow}$, and a finite spin-flip tunneling rate $\widetilde{\tau}_\mathrm{sf}$.
	As a consequence, the effective Hamiltonian is not symmetric (nor anti-symmetric) with respect to $\delta$.
	However, it does exhibit reflection symmetry $\delta\to-\delta$ and $\epsilon_\mathrm{ac}\to-\epsilon_\mathrm{ac}$.
	Under this reflection, the effective tunneling rates transform as $\widetilde{\tau}_\uparrow \rightleftharpoons \widetilde{\tau}_\downarrow$, and $\widetilde{\tau}_\mathrm{sf}\to -\widetilde{\tau}_\mathrm{sf}$ for an odd resonance $E_z=(2k+1)\omega,k\in\mathds{Z}$, as discussed in \cref{sec:DQD_global_OME}.
	
	\section*{References}
	\bibliographystyle{iopart-num}
	\bibliography{references}

\end{document}